\documentclass[aps,prd,nofootinbib,showkeys,preprint,floatfix]{revtex4}

\usepackage{amssymb}
\usepackage{amsmath}
\usepackage{amsfonts}
\usepackage{graphicx}
\usepackage{color}
\usepackage{xspace}
\usepackage{ulem}
\usepackage{lscape}
\usepackage{ wasysym }
\usepackage[toc,page]{appendix}

\usepackage{multirow,rotating}

\def\gsim{\raise0.3ex\hbox{$\;>$\kern-0.75em\raise-1.1ex\hbox{$\sim\;$}}}
\def\lsim{\raise0.3ex\hbox{$\;<$\kern-0.75em\raise-1.1ex\hbox{$\sim\;$}}}

\def\znbb{0\nu\beta\beta}

\def\rp{$R_p\hspace{-1em}/\ \ $}
\def\rpt{$R_p\hspace{-0.87em}/\ \ $}

\newcommand{\ba}[1]{\begin{eqnarray} \label{(#1)}}
\newcommand{\ea}{\end{eqnarray}}

\newcommand{\AddrAHEP}{
  {\it AHEP Group, Instituto de F\'{\i}sica Corpuscular --
    C.S.I.C./Universitat de Val{\`e}ncia \\
    Edificio de Institutos de Paterna, Apartado 22085,
  E--46071 Val{\`e}ncia, Spain}}

\newcommand{\AddrUFSM}{
Universidad T\'ecnica Federico Santa Mar\'\i a, \\ 
Centro-Cient\'\i fico-Tecnol\'{o}gico de Valpara\'\i so, \\ 
Casilla 110-V, Valpara\'\i so,  Chile}

\def\gsim{\raise0.3ex\hbox{$\;>$\kern-0.75em\raise-1.1ex\hbox{$\sim\;$}}}
\def\lsim{\raise0.3ex\hbox{$\;<$\kern-0.75em\raise-1.1ex\hbox{$\sim\;$}}}

\begin{document}

\preprint{IFIC/16-83}  

\title{QCD-improved limits from neutrinoless double beta decay}

\author{C. Arbel\'aez}\email{carolina.arbelaez@usm.cl}\affiliation{\AddrUFSM}
\author{M. Gonz\'alez} \email{marcela.gonzalezp@usm.cl}\affiliation{\AddrUFSM}
\author{M. Hirsch} \email{mahirsch@ific.uv.es}\affiliation{\AddrAHEP}
\author{S.G. Kovalenko}\email{Sergey.Kovalenko@usm.cl}\affiliation{\AddrUFSM}



\begin{abstract}
We analyze the impact of QCD corrections on limits derived from
neutrinoless double beta decay ($\znbb$).  As demonstrated previously,
the effect of the color-mismatch arising from loops with gluons
linking the quarks from different color-singlet currents participating
in the effective operators has a dramatic impact on the predictions
for some particular Wilson coefficients. Here, we consider all
possible contributions from heavy particle exchange, i.e. the
so-called short-range mechanism of $\znbb$ decay. All high-scale
models (HSM) in this class match at some scale around a $\sim$ few TeV
with the corresponding effective theory, containing a certain set of
effective dimension-9 operators.  Many of these HSM receive
contributions from more than one of the basic operators and we
calculate limits on these models using the latest experimental data.
We also show with one non-trivial example, how to derive limits on
more complicated models, in which many different Feynman diagrams
contribute to $\znbb$ decay, using our general method.

\end{abstract}

\maketitle

%%%%%%%%%%%%%%%%%%%%%%%%%%%%%%%%%%%%%%%%%%%%%%%%%%%%%%%%%%%%%%%%%%%%%%
%\tableofcontents

\section{Introduction}
\label{sec:introduction}

Lepton Number Violation (LNV) appears in many extensions of the
Standard Model (SM). If LNV exists, it could be the explanation for
the smallness of the observed neutrino masses and maybe even the
baryon asymmetry of the universe \cite{Fukugita:1986hr}.  Neutrinoless
double beta decay ($\znbb$) is widely credited as the most promising
probe for LNV from the view point of experimental observability.
Consequently, $\znbb$-decay has been studied in great detail, both
from theoretical and experimental points of view. \footnote{For
  reviews of particle physics aspects of $\znbb$ see for instance
  refs.~\cite{Deppisch:2012nb,Rodejohann:2011mu} and for recent
  calculations of nuclear matrix elements
  refs.~\cite{Faessler:2012ku,Simkovic:2013kna}.}

A number of experiments are currently searching for $\znbb$-decay
\cite{Agostini:2013mzu,Albert:2014awa,Shimizu:2014xxx,KamLAND-Zen:2016pfg,GERDAII:2016}
with the negative results setting lower bounds on the $\znbb$-half-life
$T_{1/2}^{0\nu}$. Currently the best bounds are
\begin{eqnarray}\label{eq:T0nu-1}
\mbox{KamLAND-Zen} \ 
\mbox{\cite{KamLAND-Zen:2016pfg}}
&:&  T^{0\nu}_{1/2}({}^{136}{\rm Xe}) = 1.07\times 10^{26} \ {\rm ys} 
\ (90\% {\rm C.L.}),\\
\label{eq:T0nu-2} 
\mbox{GERDA Phase-II }\ \mbox{\cite{GERDAII:2016}}&:& 
T^{0\nu}_{1/2}({}^{76}{\rm Ge})\  = 5.2\ \, \times 10^{25} \ {\rm ys}
\  (90\% {\rm C.L.}).
\end{eqnarray}
Sensitivities in excess of $T^{0\nu}_{1/2}\gsim 10^{27}$ ys in
experiments using ${}^{136}$Xe \cite{EXO-200:2013rm} and ${}^{76}$Ge
\cite{Abt:2004yk,Guiseppe:2011me} are expected in the future.

\begin{figure}[t]
%\centering
\includegraphics[width=0.43\linewidth]{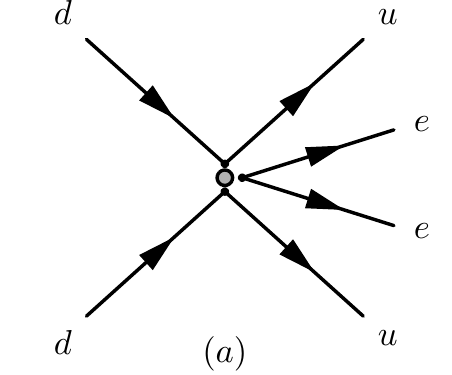}
\hskip10mm\includegraphics[width=0.37\linewidth]{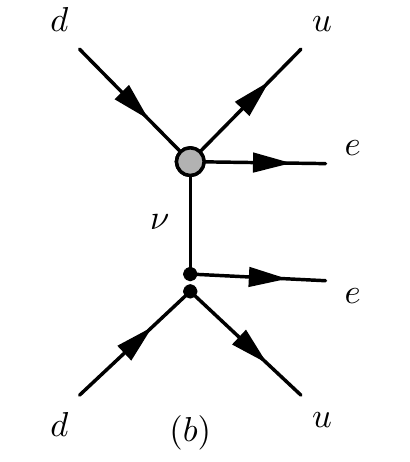}
\caption{Short-range mechanism (SRM), to the left, and long-range
  mechanism (LRM) to the right. The grey blobs indicate effective
  vertices originating from heavy particle-exchange.
}
\label{fig:srmlrm}
\end{figure}

Contributions to $\znbb$-decay can be classified as either long-range (LRM) 
\cite{Pas:1999fc} or short-range mechanisms (SRM)
\cite{Pas:2000vn}, depending on whether all of the virtually exchanged
particles are heavy or not, see Fig.~\ref{fig:srmlrm}. For 
the short-range mechanisms,
SRM, the experimental limits imply typical masses of heavy intermediate
particles and an LNV scale $\Lambda_{LNV}$ in the ballpark of (a few)
TeV.  Therefore, the LHC could possibly provide a cross-check whether
or not these contributions can be dominant in $\znbb$-decay 
\cite{Helo:2013dla,Helo:2013ika,Helo:2013esa,Helo:2015ffa,Gonzales:2016krw}.

Naturally, for a realistic comparison of the sensitivities of
$\znbb$-decay with the LHC the theoretical calculations must be made
as reliable as possible, which is particularly demanding for
$\znbb$-decay.  One well-known source of difficulties in this case are
uncertainties in the Nuclear Matrix Elements (NME), which spread by a
factor of typically $\sim$ 2 comparing different
calculations. Improving the predictions for $\znbb$-NMEs is a serious
challenge for nuclear structure theory, which is going to take time
and significant efforts. On the other hand, recently it has been
pointed out that one important effect has so far been missing in the
theoretical treatment of $\znbb$-decay
\cite{Mahajan:2014twa,Gonzalez:2015ady}: QCD corrections.  This
effect, being perturbative, is much better controllable theoretically
than the essentially non-perturbative (in quantum field theory sense)
physics involved in the NME-calculations.  As explained in
Ref.~\cite{Gonzalez:2015ady} gluon exchange diagrams can lead to the
so called ``color-mismatch'' in the products of the color-singlet
quark currents giving rise to an appreciable mixing between different
$\znbb$-effective operators. The vastly differing numerical values of
NMEs for different operators then result in dramatic changes of the
limits on the Wilson coefficients of some particular operators. This
feature is pertinent to the SRMs of $\znbb$-decay.  It has been
recently demonstrated in Ref.~\cite{Arbelaez:2016zlt} that the
color-mismatch effect is absent in the case of the LRMs of
$\znbb$-decay and, therefore, for this class of mechanisms the QCD
corrections are not so crucial.

For such a fairly low-energy process as $\znbb$-decay an effective
operator description is adequate for calculating the decay rate.  This
rather straightforward observation forms the framework of the original
papers \cite{Pas:1999fc,Pas:2000vn} where the basis of the effective
$\znbb$-decay $d=9$ operators was introduced and generic formulas for
the $\znbb$-decay half-live were derived.  More recently this approach
has been developed in Refs.~\cite{Gonzalez:2015ady,Arbelaez:2016zlt}
where we derived the QCD-corrected $\znbb$-decay half-life formulas
for both SRM and LRM
\footnote{In Ref.~\cite{Peng:2015haa} the QCD corrections were taken
  into account to the pion-exchange mechanism \cite{Faessler:1997db}
  of $\znbb$-decay.}.
From the more fundamental high-energy point of view, however, contributions to 
$\znbb$-decay originate from some renormalizable LNV extension of the
SM, i.e. high-scale models (HSM), whose parameters are the
couplings and masses of experimentally yet unknown particles.  

A list of all possible HSMs representing UV completions of the
above-mentioned $\znbb$-decay $d=9$ operators was given in
Ref.~\cite{Bonnet:2012kh} and from that paper, in principle, all the
HSMs contributing to $\znbb$-decay via the short-range mechanism can
be found.  The purpose of our present paper is to provide a bridge
between these two descriptions -- in terms of the effective operators
and the HSMs -- taking into account the effect of the above-mentioned
QCD corrections. Upper limits derived on the Wilson coefficients of
the $\znbb$-decay effective operators (low energy approach)
\cite{Gonzalez:2015ady} can be converted into lower limits on the mass
scales of the HSMs listed in Ref.~\cite{Bonnet:2012kh} and we provide
tables of these limits, using updated experimental lower bounds on the
$\znbb$-decay half-life, for all ``elementary'' HSMs (see section
\ref{sec:short-range-limits}).  While these ``translation rules'' can
be applied in a rather straightforward manner to any particular HSM,
for which only one Feynman diagram contributes significantly, there
are many example models in the literature where this is not the
case. In the presence of more than one significant diagram a careful
examination of their contributions to different operators is required
for arriving at the correct answer.  We will discuss one particular
example -- R-parity violating supersymmetry -- in some detail, to
demonstrate the usefulness of our approach.

This paper is organized as follows.  In Sec. \ref{sec:short-range} we
start by recalling the definitions for the QCD corrected half-life
formula for $\znbb$-decay. This section summarizes the results of
Ref.~\cite{Gonzalez:2015ady}. We then derive limits on ``elementary''
HSMs contributing to the short-range $\znbb$-decay mechanism in
section \ref{sec:short-range-limits}.  In Sec. \ref{sec:RPV} we
discuss how to derive limits in our approach on more complicated
HSMs. As already mentioned, we choose the well-known example of
R-parity violating SUSY.  We conclude with a discussion of our results
in Sec.~\ref{sec:Discussion}. Some more technical aspects of the
calculation are delegated to an Appendix.

%%%%%%%%%%%%%%%%%%%%%%%%%%%%%%%%%%%%%%%%%%%%%%%%%%%%%%%%%%%%%%%%%%%%%%

\section{QCD running of  Short-range Mechanisms}
\label{sec:short-range}

\begin{figure}[t]
\includegraphics[width=0.3\linewidth]{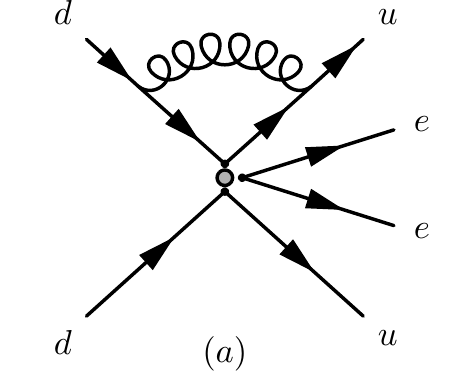}
\includegraphics[width=0.3\textwidth]{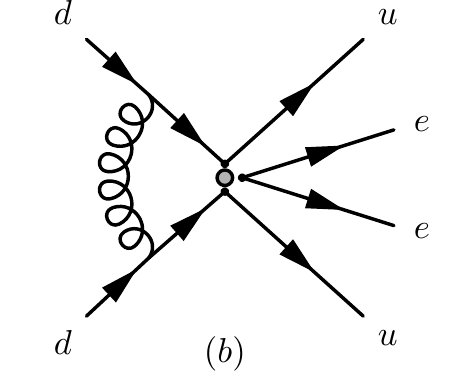}
\includegraphics[width=0.3\textwidth]{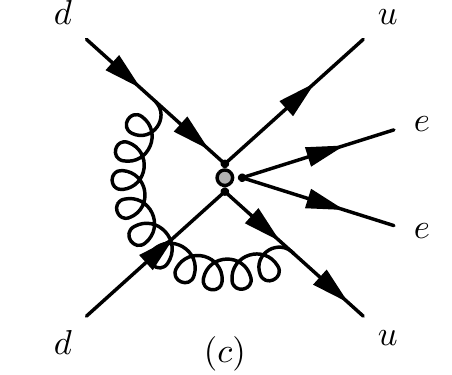}
\caption{One-loop QCD corrections to the short range mechanisms of 
$0\nu\beta\beta$ decay in the effective theory.}
\label{fig:qcdc}
\end{figure}

The contribution of a HSM to $\znbb$-decay via heavy particle exchange
we call the short-range mechanism (SRM), already mentioned in
Introduction.  After integrating out the heavy degrees of freedom of a
mass $\sim M_{I}$ at an energy-scale $\mu< M_{I}$, all the HSMs of the
SRM category can be represented by the effective Lagrangian
\cite{Pas:2000vn,Gonzalez:2015ady}
\begin{eqnarray}\label{eq:LagGen}
{\cal L}^{0\nu\beta\beta}_{\rm eff} = \frac{G_F^2}{2 m_p} \,
              \sum_{i, XY} C_{i}^{XY}(\mu)\cdot \mathcal{O}^{XY}_{i}(\mu),
\end{eqnarray}
with the complete set of  dimension-9 $\znbb$-operators
\begin{eqnarray}
\label{eq:OperBasis-1}
\mathcal{O}^{XY}_{1}&=& 4 ({\bar u}P_{X}d) ({\bar u}P_{Y}d) \ j,\\
\label{eq:OperBasis-2}
\mathcal{O}^{XX}_{2}&=& 4 ({\bar u}\sigma^{\mu\nu}P_{X}d)
                         ({\bar u}\sigma_{\mu\nu}P_{X}d) \ j,\\
\label{eq:OperBasis-3}
\mathcal{O}^{XY}_{3}&=& 4 ({\bar u}\gamma^{\mu}P_{X}d) 
                        ({\bar u}\gamma_{\mu}P_{Y}d) \  j,\\
\label{eq:OperBasis-4}
\mathcal{O}^{XY}_{4}&=& 4 ({\bar u}\gamma^{\mu}P_{X}d) 
                         ({\bar u}\sigma_{\mu\nu}P_{Y}d) \ j^{\nu},\\
\label{eq:OperBasis-5}
\mathcal{O}^{XY}_{5}&=& 4 ({\bar u}\gamma^{\mu}P_{X}d) ({\bar u}P_{Y}d) \ j_{\mu},
\end{eqnarray}
where $X,Y = L,R$ and the leptonic currents are 
\begin{eqnarray}\label{eq:Curr}
j = {\bar e}(1\pm \gamma_{5})e^c \, , \quad j_{\mu} = {\bar e}\gamma_{\mu}\gamma_{5} e^c .
\end{eqnarray}
The Wilson coefficients $C_{i}^{XY}$ can be expressed in terms of the
parameters of a particular HSM at a scale $\Lambda \sim M_{I}$, called
``matching scale''.  Note that some of $C_{i}(\Lambda)$ may vanish.
In order to make contact with $\znbb$-decay one needs to estimate
$C_{i}$ at a scale $\mu_{0}$ close to the typical $\znbb$-energy
scale. The QCD corrections, such as shown in Fig.~\ref{fig:qcdc}, lead
to running of the coefficients between the matching $\Lambda$ and
$\mu_{0}$ scales.

While the QCD-running is only logarithmic, it mixes different
operators (or equivalently Wilson coefficients) from the list
(\ref{eq:OperBasis-1})-(\ref{eq:OperBasis-5}). Because of the vast
difference of the NMEs of some operators, this effect results in a
dramatic impact on the prediction of some HSM for $\znbb$-decay
\cite{Gonzalez:2015ady}.  The $\znbb$-decay half-life formula, taking
into account the leading order QCD-running \cite{Gonzalez:2015ady}, reads 
\begin{eqnarray}\label{eq:T12} 
\Big[ T^{\znbb}_{1/2}\Big]^{-1} = 
G_1 \left|\sum_{i=1}^{3} \beta_{i}^{XY}(\mu_{0}, \Lambda) C^{XY}_{i}(\Lambda) \right|^2 +  
G_2 \left|\sum_{i=4}^{5} \beta_{i}^{XY}(\mu_{0}, \Lambda) C^{XY}_{i}(\Lambda)\right|^2
\end{eqnarray}
Here, $G_{1,2}$ are phase space factors \cite{Doi:1985dx,Pas:2000vn}.
The parameters $\beta_{i}^{XY}$ incorporate the QCD-running and the
NMEs of the operators in
Eqs.~(\ref{eq:OperBasis-1})-(\ref{eq:OperBasis-5}).  We show the
values of these coefficients in Table~\ref{t:betaparameters}
calculated with the NMEs from Refs.~\cite{Deppisch:2012nb}.  In
Eq.~(\ref{eq:T12}) the summation over the different chiralities $X,Y =
L,R$ is implied.  It is important to note that the Wilson coefficients
$C_{i}(\Lambda)$, entering in Eq.~(\ref{eq:T12}), are linked to the
matching scale $\Lambda$, where they are calculable in terms of the
HSM parameters, such as couplings and intermediate particle masses.

\begin{table}[h]
\begin{tabular}{|c|c|c||c|c|c|}
\hline 
\hline
Coeff.&\multicolumn{2}{c||}{Isotope}&Coeff.&\multicolumn{2}{c|}{Isotope}\\
\cline{2-3} \cline{5-6}
&${}^{76}$Ge&${}^{136}$Xe&&${}^{76}$Ge&${}^{136}$Xe\\
\hline
&&&&&\\
$\beta_{1}^{XX}$&$6.1\times 10^{3}$&$3.1\times 10^{3}$&$\beta_{4}^{XX}$&$(-5.6+0.2i)\times 10^{2}$ &$(-2.9+0.1i)\times 10^{2}$ \\
&&&&&\\
$\beta_{1}^{LR}$&$-2.5\times 10^{2}$&$-1.3\times 10^{2}$&$\beta_{4}^{LR}$&$1.2\times 10^{2}$&$6.0\times 10^{1}$\\
&&&&&\\
$\beta_{2}^{XX}$&$-4.4\times 10^{2}$&$-2.3\times 10^{2}$&$\beta_{5}^{XX}$&$(0.9 - 1.3i)\times 10^{2}$&$(4.5 - 6.6i)\times 10^{1}$\\
&&&&&\\
$\beta_{3}^{XX}$&$1.5\times 10^{2}$&$7.5\times 10^{1}$&$\beta_{5}^{LR}$&$7.7\times 10^{1}$&$3.9\times 10^{1}$\\
&&&&&\\
$\beta_{3}^{LR}$&$1.1\times 10^{2}$&$5.7\times 10^{1}$&&&\\
&&&&&\\
\hline
\hline
\end{tabular}
\caption{The coefficients $\beta_{i}\equiv \beta_{i}(\mu_{0},
  \Lambda)$ incorporating NMEs and entering the QCD corrected
  half-life formula (\ref{eq:T12}). The results are shown for the
  QCD-running between the scales $\Lambda = 1$ TeV and $\mu_{0} = 1$
  GeV.  We used NMEs from Ref.~\cite{Deppisch:2012nb}.}
\label{t:betaparameters}
\end{table}

%%%%%%%%%%%%%%%%%%%%%%%%

In Ref. \cite{Gonzalez:2015ady} we used the $\znbb$-decay half-life
formula (\ref{eq:T12}) in order to extract ``individual'' upper limits
on the Wilson coefficients $C^{XY} _{i}$ from the existing
experimental bounds on $T^{\znbb}_{1/2}$.  We employed the
conventional hypothesis that a single term dominates in
Eq.~(\ref{eq:T12}). This method disregards effects of a possible
simultaneous presence of several non-zero terms, which may partially
cancel each other or give rise to a significant enhancement.  These
effects are discussed in the next section.

\section{Limits on Short-range elementary High-scale models}
\label{sec:short-range-limits}

\begin{figure}[t]
%\centering
%\hskip-10mm\includegraphics[width=0.5\linewidth]{TopoI_SFS.eps}
%\hskip10mm\includegraphics[width=0.4\linewidth]{TopoII_SSS.eps}
\hskip-10mm\includegraphics[width=0.5\linewidth]{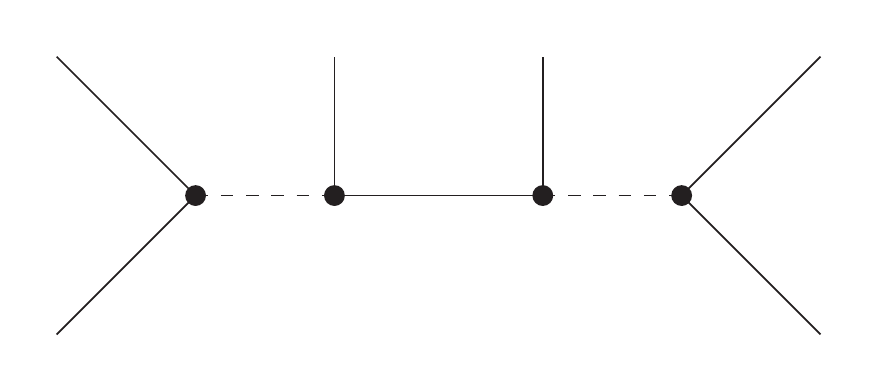}
\hskip10mm\includegraphics[width=0.4\linewidth]{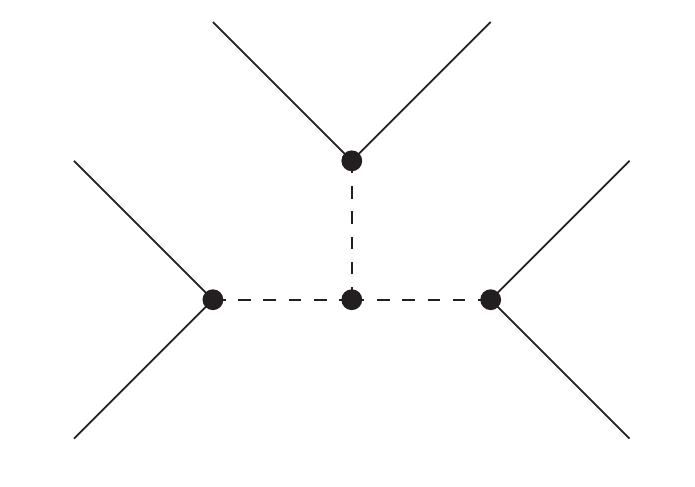}
\caption{Tree-level topologies contributing to the short-range
  mechanism of $\znbb$-decay. To the left T-I, boson-fermion-boson 
  exchange; to the right T-II, triple-boson diagrams. External lines
  stand symbolically for any permutation of ${\bar u}{\bar u}dd{\bar
    e}{\bar e}$.}
\label{fig:short}
\end{figure}

Two tree-level topologies contributing to the $\znbb$ decay amplitude
were identified in Ref.~\cite{Bonnet:2012kh}, see
Fig.~\ref{fig:short}.  Here, the outer lines of the diagrams represent
all possible permutations of the six fermions ${\bar u}{\bar u}dd{\bar
  e}{\bar e}$, which make up the $\znbb$ decay operator.  Considering
\mbox{$G_{\rm SM}=SU(3)_c\times SU(2)_L\times U(1)_Y$} invariant
vertices in these diagrams one may derive a complete list of the
$G_{\rm SM}$-assignments for the intermediate particles \mbox{(Scalar,
  Fermion, Scalar) = $(S,\Psi, S')$} and (three Scalars) = $(S, S' ,
S'')$ in the T-I and \mbox{T-II} topology diagrams, respectively. This
was done in Ref.~\cite{Bonnet:2012kh}.  Each case in this list we call
``elementary'' HSM (eHSM).  We reproduce the original list of
Ref.~\cite{Bonnet:2012kh} in Tables \ref{t:HSMIdentTI-1},
\ref{t:HSMIdentTI-2} and \ref{t:HSMIdentTII} where, for convenience of
our analysis, we collected the eHSMs in groups enumerated by \#I.
Any short-range HSM can be represented in the form of a linear
combination of several eHSMs. The HSMs considered in the literature
(for a recent review c.f. Ref.~\cite{Deppisch:2012nb}) are mainly of
this kind with the parameters (couplings, masses) of the involved
eHSMs related between each other by symmetry or other arguments.  We
will discuss one example of such non-elementary HSM -- \rp SUSY -- in
the next section and here first focus on the eHSMs.

%%%%%%%%%%%%%%%%%%%%%%%%%%%%%%%%%%%%%%%%%%%%%%%%%%%%%%%%%%%%%%%%%%%%%%

In the case of the short-range mechanism all the HSMs in the
low-energy limit are reducible to the effective Lagrangian
(\ref{eq:LagGen}).  By definition each eHSM generates, after
integrating out heavy particles, a single effective operator. It is
straightforward, although tedious, to check that all the eHSMs from
each group \#I in Tables \ref{t:HSMIdentTI-1}, \ref{t:HSMIdentTI-2}
and \ref{t:HSMIdentTII} lead to the same effective operator ${\cal
  O}^{I}$.  Projection of ${\cal O}^{I}$ on the general operator basis
${\cal O}_{i}$ in Eqs.~(\ref{eq:OperBasis-1})-(\ref{eq:OperBasis-5})
via Lorentz and color Fierz transformations (see Appendix
\ref{sec:App:Decomp}) gives rise to a linear combination of only two
basis operators
\begin{eqnarray}\label{eq:OI-basis-decomp-1}
{\cal O}^{I} =
x^{I}_{i} {\cal O}_{i}+y^{I}_{j}{\cal O}_{j}
\end{eqnarray}
with numerical coefficients $x^{I}, y^{I}$ algebraically calculable
for any particular eHSM \cite{Bonnet:2012kh}. Note that no summation
over the repeated indices $i,j$ is implied in
Eq.~\ref{eq:OI-basis-decomp-1}.  All the possible operator pairs with
the corresponding coefficients are shown in
Tables~\ref{t:HSModelsShort-range-1},
\ref{t:HSModelsShort-range-2}. Some eHSM lead to only one of the basis
operators, these are listed in Table~\ref{t:HSModelsShort-range-3}.
The values of the coefficients $x^{I}, y^{I}$ given in these Tables
are useful as an additional identifier of the eHSMs as well as for
recalculation of the experimental limits on the parameters of the
eHSMs, also given in these Tables, with the NMEs and the experimental
$\znbb$-decay half-life bounds different from those we used here.

We derived these limits in the following way.  The effective Lagrangian for
any  \#I  eHSM at the matching scale $\Lambda$ can be expressed, 
taking into account (\ref{eq:OI-basis-decomp-1}), as
\begin{eqnarray}\label{eq:Eff-EHSM-1}
&&{\cal L}_{I} = \frac{G_F^2}{2 m_p} \,
          C_{I}(\Lambda)\cdot \mathcal{O}^{I}(\Lambda) = \frac{G_F^2}{2 m_p} \,
             C_{I}(\Lambda)\cdot \left(x^{I}_{i} \mathcal{O}_{i}(\Lambda) + 
             y^{I}_{j} \mathcal{O}_{j}(\Lambda)\right).
\end{eqnarray}
The half-life formula (\ref{eq:T12}), used to constrain a concrete
eHSM, is reduced to
\begin{eqnarray}\label{eq:T-mod-1}
T_{1/2}^{-1}=G_{K_{I}} \left|C_{I}(\Lambda)\right|^{2} 
\left|\beta_{i}(\mu_{0}, \Lambda) x^{I}_{i} + 
\beta_{j}(\mu_{0}, \Lambda) y^{I}_{j} \right|^{2}.
\end{eqnarray}
Here $K_{I} = 1$ or $2$ for $i,j \in \{1,2,3\}$ or for $i,j \in
\{4,5\}$, respectively.

Using the current experimental $\znbb$-decay half-life lower bounds
(\ref{eq:T0nu-1}), (\ref{eq:T0nu-2}) we derive from
Eq.~(\ref{eq:T-mod-1}) upper limits on the Wilson coefficients
$C_{I}(\Lambda= 1 {\rm TeV})\leq C_{I}^{exp}$.  These limits are shown
in Tables~\ref{t:HSModelsShort-range-1},
\ref{t:HSModelsShort-range-2}, \ref{t:HSModelsShort-range-3}.  For a
more direct comparison of the $\znbb$-decay limits with the
sensitivity of an accelerator experiment, such as the LHC, it is
instructive to convert these limits into limits on the scale $M_{I}$
of the masses of the intermediate heavy particles mediating the
contribution of \#I eHSMs to $\znbb$-decay.  Denoting the
dimensionless couplings in the T-I diagram from Figs.~\ref{fig:short}
with $\lambda_{1,2,3,4}$ and letting all the intermediate particle
masses be of the order of the same scale $M_{I}$ we can give an
estimation
\begin{eqnarray}\label{eq:mass-scale-1}
\frac{G_F^2}{2 m_p} \,
             C_{I}(\Lambda)= \frac{\lambda_{1}\lambda_{2}\lambda_{3}\lambda_{4}}{M_{I}^{5}}
\end{eqnarray}
for the overall coefficient in Eq.~(\ref{eq:Eff-EHSM-1}) .  Then we find
lower limits for the typical mass scale at which a particular eHSM
contributes to the short-range mechanism:
\begin{eqnarray}\label{eq:mass-lim-1} 
M_{I} \geq  \lambda_{eff}^{4/5}\left(\frac{2 m_{p}}{C_{I}^{exp} G_{F}^{2}} \right)^{1/5} =
\lambda_{eff}^{4/5} \, \overline{M}_{I}^{exp},
\end{eqnarray}
where we introduced for convenience $\lambda_{eff} =
(\lambda_{1}\lambda_{2}\lambda_{3}\lambda_{4})^{1/4}$ and
$C_{I}^{exp}$ are the previously derived upper limits on
$C_{I}(\Lambda)\leq C_{I}^{exp}$. We show these limits in
Tables~\ref{t:HSModelsShort-range-1}-\ref{t:HSModelsShort-range-3} for
completeness. Note that for T-II diagrams in Fig.~\ref{fig:short}, the
triple-scalar coupling has dimension of mass. Nevertheless, we can
apply the same limits, as in the case of T-I, assuming this coupling
to be of order $\mu = \lambda_{eff}M_{I}$.

%%%%GOOD TABLES
%
\begin{table}[h]
\centering
%\scalebox{0.5}
\begin{tabular}{c|c|c|c|c|c|c}
%\scalebox{0.3}
\hline\hline
eHSM&Effective operator &$x, y$&\multicolumn{2}{c|}{$C^{exp}_{I}$}&\multicolumn{2}{c}{
$\overline{M}_{I}^{exp}$ TeV}\\ \cline{4-7}
\#I& decomposition& &$^{76}$Ge &$^{136}$Xe& \ $^{76}$Ge \ \ \ \ &$^{136}$Xe\\
\hline 
& $\mathcal{O}^{I} = x \ \mathcal{O}_1^{XX}+ y \ \mathcal{O}_2^{XX}$ &&&&&\\

1&&$-\frac{5}{24},-\frac{1}{32}$ & $1.4\times 10^{-9}$ & $6.9\times 10^{-10}$&$6.3$&$7.3$\\&&&&&\\

2&&$\frac{1}{32},\frac{1}{128}$& $9.2\times 10^{-9}$ &$4.6\times 10^{-9}$ &$4.3$ &$5.0$\\&&&&&\\

3&&$-\frac{7}{48},-\frac{1}{192}$& $2.0\times 10^{-9}$ &$9.7\times 10^{-10}$ & $5.9$& $6.8$\\&&&&&\\

4&&$\frac{1}{32},-\frac{1}{128}$& $8.9\times 10^{-9}$  &$4.4\times 10^{-9}$  &$4.4$ &$5.0$\\&&&&&\\

5&&$\frac{1}{24},-\frac{1}{96}$ & $6.7\times 10^{-9}$ & $3.3\times 10^{-9}$ &$4.6$&$5.3$\\&&&&&\\

6&&$\frac{3}{32},\frac{1}{128}$& $3.0\times 10^{-9}$ &$1.5\times 10^{-9}$ &$5.4$&$6.2$\\&&&&&\\

\hline 
&$\mathcal{O}^{I} = x \mathcal{O}_1^{LR,RL}+ y \mathcal{O}_3^{LR,RL}$ &&&&&\\
7&&$-\frac{1}{12},-\frac{1}{8}$&  $2.4\times 10^{-7}$ &$1.2\times 10^{-7}$ &$2.3$&$2.6$\\&&&&&\\
8&&$-\frac{1}{8},-\frac{1}{48}$& $6.0\times 10^{-8}$  &$2.9\times 10^{-8}$ &$3.0 $& $3.4$ \\&&&&&\\
9&&$\frac{1}{16},\frac{1}{32}$& $1.4\times 10^{-7}$  &$7.0\times 10^{-8} $ &$2.5 $&$2.9$\\&&&&&\\

10&&$-\frac{1}{32},\frac{1}{64}$&$1.8\times 10^{-7}$& $8.9\times 10^{-8} $ &$2.4$&$2.7$\\&&&&&\\

\hline\hline
\end{tabular}
\caption{Decomposition in the basis operators
  (\ref{eq:OperBasis-1})-(\ref{eq:OperBasis-5}) of the effective
  operators $\mathcal{O}^{I}$ representing low-energy limits of the
  eHSMs of the group \# I specified in
  Tables~\ref{t:HSMIdentTI-1}-\ref{t:HSMIdentTII}.  Experimental
  limits on the Wilson Coefficients $C_{I}(\Lambda= 1 {\rm TeV})\leq
  C_{I}^{exp}$ of these operators and their characteristic scales
  $M_{I}\geq \lambda_{eff}^{4/5} \cdot \overline{M}_{I}^{exp}$ (for
  the definitions see Eqs.~(\ref{eq:Eff-EHSM-1}),
  (\ref{eq:mass-lim-1})) are derived from the current $\znbb$ bounds
  (\ref{eq:T0nu-1}), (\ref{eq:T0nu-2}).}
\label{t:HSModelsShort-range-1}
\end{table}

\begin{table}[h]
\centering
%\scalebox{0.5}
\begin{tabular}{c|c|c|c|c|c|c}
%\scalebox{0.3}
\hline\hline
eHSM&Effective operator&$\alpha, \beta$&\multicolumn{2}{c|}{$C_{I}^{exp}$}&\multicolumn{2}{c}{$\overline{M}_{I}^{exp}$ TeV}\\ \cline{4-7}
\#I& decomposition & &$^{76}$Ge &$^{136}$Xe& \ $^{76}$Ge \ \ \ \ &$^{136}$Xe\\
\hline 
& $\mathcal{O}^{I} = x \mathcal{O}_4^{XX}+ y \mathcal{O}_5^{XX}$ &&&&&\\
11&&$-\frac{1}{16i},-\frac{5}{48}$& $6.9\times 10^{-8}$ &$3.5\times 10^{-8}$ &$2.9 $&$3.3$ \\&&&&&\\

12&&$-\frac{1}{32i},-\frac{1}{32}$& $1.2\times 10^{-7}$  & $6.0\times 10^{-8}$&$2.6$&$3.0$\\&&&&&\\

13&&$\frac{1}{32i},-\frac{1}{32}$& $7.6\times 10^{-8}$  & $3.9\times 10^{-8}$ &$2.8$&$3.2$\\&&&&&\\

14&&$\frac{1}{48i},\frac{7}{48}$& $1.1\times 10^{-7}$ &$5.5\times 10^{-8}$ &$2.7 $&$3.0$\\&&&&&\\
15&&$-\frac{1}{32i},-\frac{3}{32}$& $1.6\times 10^{-7}$ &$8.1\times 10^{-8}$ & $2.5 $&$2.8$\\&&&&&\\
16&&$\frac{1}{24i},-\frac{1}{24}$& $5.7\times 10^{-8}$ &$2.9\times 10^{-8}$ &$3.0 $ &$3.4$\\&&&&&\\

\hline
& $\mathcal{O}^{I} =  x\mathcal{O}_4^{LR,RL}+ y\mathcal{O}_5^{LR,RL}$ &&&&&\\

17&&$\frac{1}{16i},-\frac{5}{48}$&  $1.5\times 10^{-7}$ &$7.8\times 10^{-8}$&$2.5 $& $2.8$\\&&&&&\\

18&&$\frac{1}{32i},-\frac{1}{32}$&  $3.8\times 10^{-7}$ &$1.9\times 10^{-7}$ &$2.1 $&$2.4$\\&&&&&\\

19&&$-\frac{1}{48i},\frac{7}{48}$&  $1.4\times 10^{-7}$ &$7.4\times 10^{-8}$ &$2.5 $&$2.9$\\&&&&&\\

20&&$\frac{1}{32i},-\frac{3}{32}$& $2.0\times 10^{-7}$  &$1.1\times 10^{-7}$ &$2.3 $&$2.7$\\&&&&&\\

21&&$-\frac{1}{32i},-\frac{1}{32}$& $3.7\times 10^{-7}$   & $1.9\times 10^{-7}$  &$2.1$&$2.4$\\&&&&&\\

22&&$-\frac{1}{24i},-\frac{1}{24}$& $2.8\times 10^{-7}$  & $1.4\times 10^{-7}$  &$2.2$ &$2.5$ \\&&&&&\\

\hline\hline
\end{tabular}
\caption{Continuation of Table~\ref{t:HSModelsShort-range-1}.}
\label{t:HSModelsShort-range-2}
\end{table}

\begin{table}[h]
\centering
%\scalebox{0.5}
%\begin{tabular}{c|c|c|c|c|c|c}
\begin{tabular}{c|c|c|c|c|c}
%\scalebox{0.3}
\hline\hline
eHSM&Effective operator
%&$\alpha$
&\multicolumn{2}{c|}{$C_{I}^{exp}$}&\multicolumn{2}{c}{
$\overline{M}^{exp}$ TeV}\\ \cline{3-6}
\#I& decomposition& $^{76}$Ge &$^{136}$Xe& \ $^{76}$Ge \ \ \ \ &$^{136}$Xe\\
\hline 
%
%& $ \mathcal{O}^{I} = \frac{1}{8}\mathcal{O}_1^{XX}$ &&&&&\\

23&$ \mathcal{O}^{I} = \frac{1}{8}\mathcal{O}_1^{XX}$
%&$\frac{1}{8}$
& $2.3\times 10^{-9}$  & $1.1\times 10^{-9}$& $5.7$&$6.6$\\&&&&\\

\hline

%& $ \mathcal{O}^{I} =  \frac{1}{8} \mathcal{O}_1^{LR,RL}$  &&&&&\\

24&$ \mathcal{O}^{I} =  \frac{1}{8} \mathcal{O}_1^{LR,RL}$ 
%&$\frac{1}{8}$
&  $5.2\times 10^{-8}$ & $2.7\times 10^{-8}$ &$3.0$&$3.5$\\
&&&&\\

%& $\mathcal{O}^{I} =  \frac{1}{32}  \mathcal{O}_3^{LR,RL}$  &&&&&\\
\hline

25&$\mathcal{O}^{I} =  \frac{1}{32}  \mathcal{O}_3^{LR,RL}$
%&$\frac{1}{32}$
&  $5.0\times 10^{-7}$ & $2.5\times 10^{-7}$ &$1.9$&$2.2$\\
&&&&\\

%28&&$-\frac{1}{48}$&  $7.57\times 10^{-7}$ & $3.75\times 10^{-7}$ &$1.79$&$2.06$\\&&&&&\\

\hline 
%&$\mathcal{O}^{I} =  \frac{1}{16}  \mathcal{O}_5^{XX}$ &&&&&\\

26&$\mathcal{O}^{I} =  \frac{1}{16}  \mathcal{O}_5^{XX}$
%&$\frac{1}{16}$
& $1.7\times 10^{-7}$ & $8.5\times 10^{-8}$ &$2.4$&$2.8$ \\
&&&&\\
\hline 
%&$\mathcal{O}^{I} =  \frac{1}{16} \mathcal{O}_5^{LR}$ &&&&&\\

27&$\mathcal{O}^{I} =  \frac{1}{16} \mathcal{O}_5^{LR}$ 
%&$\frac{1}{16}$
& $3.4\times 10^{-7}$ & $1.8\times 10^{-7}$ &$2.1$&$2.4$\\
&&&&\\
\hline 
%&$\mathcal{O}^{I} =  \frac{1}{16}  \mathcal{O}_3^{XX}$ &&&&&\\

28&$\mathcal{O}^{I} =  \frac{1}{16}  \mathcal{O}_3^{XX}$ 
%&$\frac{1}{16}$
&$1.9\times 10^{-7}$ & $9.5\times 10^{-8}$ &$2.4$&$2.7$\\
&&&&\\
\hline\hline
\end{tabular}
\caption{The same as in Table~\ref{t:HSModelsShort-range-1}, but for
  eHSMs decomposing in only one of the basis operators
  Eqs.~(\ref{eq:OperBasis-1})-(\ref{eq:OperBasis-5}).}
  \label{t:HSModelsShort-range-3}
\end{table}

Closing this section we emphasize once more the importance of the QCD
corrections for some particular short-range HSMs.  The largest impact
is found for models containing the operator
$\mathcal{O}^{XX}_{1}$. For example, from Table
\ref{t:HSModelsShort-range-3} one finds a lower limit on
$\Lambda_{LNV}\sim M_{I}$ of the order $\Lambda_{LNV} \gsim 6.6$
TeV. The corresponding number without QCD corrections would be
\mbox{$\Lambda_{LNV} \gsim 1.8$ TeV}. For a detailed comparison of
the limits with and without the QCD corrections we refer the reader to
Ref.~\cite{Gonzalez:2015ady}.

\section{A non-trivial example: \rpt SUSY}
\label{sec:RPV}

In the previous section we derived limits on eHSMs. Here, we discuss
how to derive limits on models, which contribute with more than one
diagram of the type T-I and/or T-II in Fig.~\ref{fig:short} to the
short-range amplitude of $\znbb$-decay. In terms of the previous
section these HSMs are linear combinations of certain eHSMs from the
list given in Tables~\ref{t:HSMIdentTI-1}-\ref{t:HSMIdentTII}.  The
example we have chosen is the well-known case of R-parity violating
supersymmetry ($R_p \hspace{-1em}/\;\:$SUSY).

\begin{figure}[t]
\includegraphics[width=0.32\linewidth]{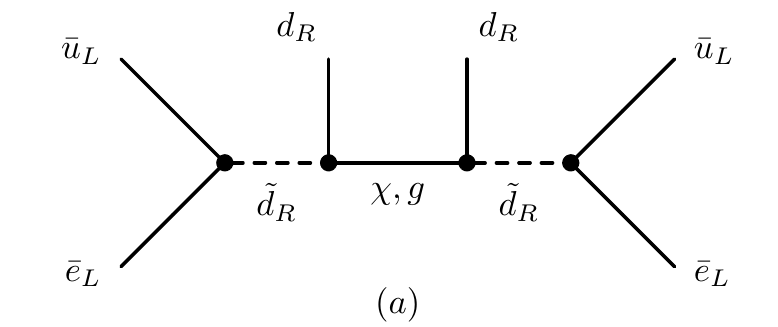}
\includegraphics[width=0.32\linewidth]{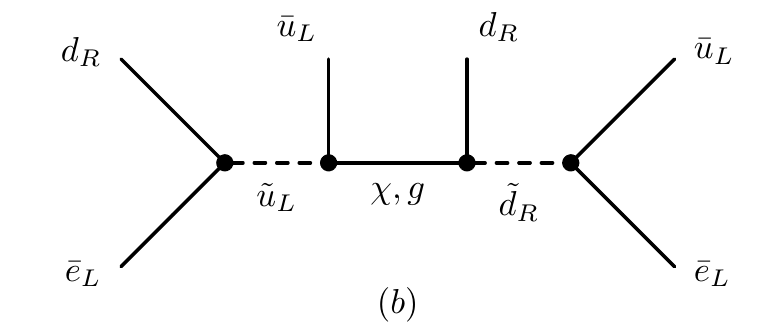}
\includegraphics[width=0.32\linewidth]{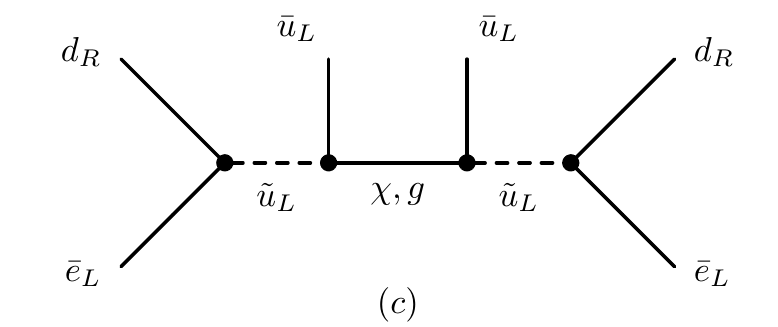}
\\
\vskip3mm
\includegraphics[width=0.32\linewidth]{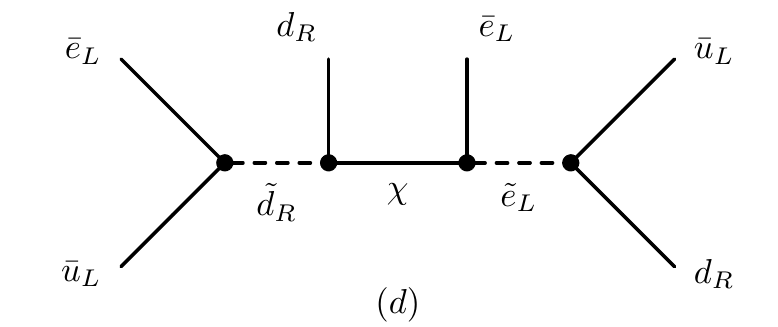}
\includegraphics[width=0.32\linewidth]{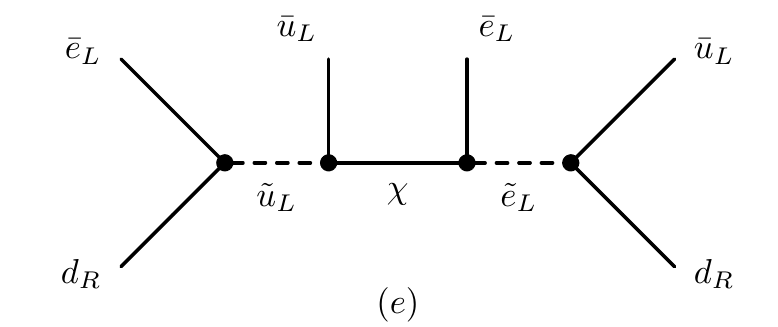}
\includegraphics[width=0.32\linewidth]{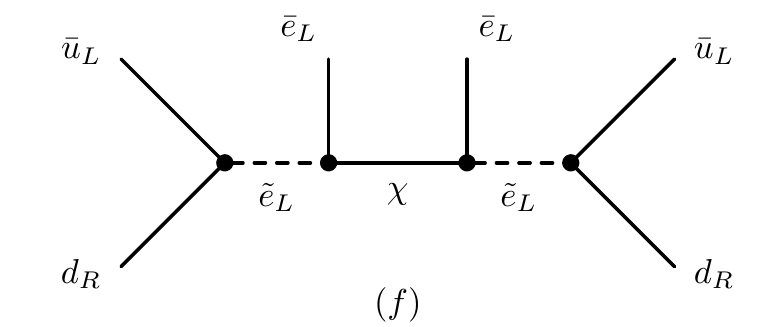}

\caption{The six different Feynman diagrams in R-parity violating 
supersymmetry that contribute to $\znbb$ decay.}
\label{fig:rpv}
\end{figure}

It provides LNV vertices with $\Delta L =1$ from the
superpotential. Importantly, in this model there are the gluino
($\tilde{g}$) and neutralino ($\chi$) Majorana mass terms, originating
from the soft SUSY breaking sector. Then $R_p \hspace{-1em}/\;\:$SUSY
can contribute to a $\Delta L= 2$ process, such as $\znbb$-decay, via
the short-range mechanism \cite{Mohapatra:1986su,Hirsch:1995ek} given
by Feynman diagrams of the topology T-I in Fig.~\ref{fig:short} with
two $\Delta L =1$ vertices, two squarks ($\tilde{q}$) or two
selectrons ($\tilde{e}$) and a $\tilde{g}$ or $\chi$ in the
intermediate state.  There are in total three gluino plus six
neutralino diagrams Ref.~\cite{Hirsch:1995ek}, see Fig.~\ref{fig:rpv}.
It is worth noting that the gluino exchange is known to give the
dominant contribution in significant parts of the minimal
$R_p \hspace{-1em}/\;\:$SUSY parameter space \cite{Hirsch:1995ek}.
Below we consider the gluino $\tilde{g}$ and the neutralino
$\chi$-exchange contributions separately, as if they were uncorrelated
sectors.  To make contact with our general method, we first identify
the transformation properties of the internal SUSY particles,
appearing in the diagram T-I in Fig.~\ref{fig:rpv}. The scalars
${\tilde u}_L$ and ${\tilde e}_L$ are members of the $SU(2)_L$
doublets ${\tilde Q}_L$ and ${\tilde L}$, respectively. The SM gauge
group assignments of the internal states of the diagrams are then
given as: ${\tilde Q_L}=S_{3,2,1/6}$, ${\tilde d_R}=S_{3,1,-1/3}$,
${\tilde L}=S_{1,2,1/2}$, ${\tilde g}=\psi_{8,1,0}$. For the
simplicity we consider the case of Bino-dominant lightest neutralino,
then $\chi=\psi_{1,1,0}$.

From Tables~\ref{t:HSMIdentTI-1}-\ref{t:HSMIdentTII} we then identify
the operator combination corresponding to each diagram.  For the
gluino diagrams this results in: diagram (a) corresponds to eHSM
$\#5$, (b) to $\#3$ and (c) again to $\#5$. The neutralino diagrams
are: (a) and (c) correspond to $\#4$, (b) to $\#2$, while diagrams
(d)-(f) can be identified with $\#23$. Note that (d) and (e) come with
an additional factor of $-\frac{1}{2}$, see
Table~\ref{t:HSMIdentTI-3}.

From these considerations, we can re-construct the corresponding
effective Lagrangians in the basis
(\ref{eq:OperBasis-1})-(\ref{eq:OperBasis-5}).  In this way we find:\\
\underline{$\tilde{g}$-{\it exchange contribution}}:
\begin{eqnarray}\label{eq:Gluino-1}
{\cal L}_{eff}^{\tilde{g}} &=&
\frac{G_F^2}{2 m_{_p}}
\left(
  C_{{\tilde g}a}\ {\cal O}_{a} 
+ C_{{\tilde g}b}\ {\cal O}_{b} 
+ C_{{\tilde g}c}\ {\cal O}_{c} \right)=\\
\nonumber
&=& \frac{G_F^2}{2 m_{_p}} \frac{1}{48}
\left[\left(2 C_{{\tilde g}a} 
+ 2 C_{{\tilde g}c} 
- 7 C_{{\tilde g}b}\right) {\cal O}^{RR}_{1} - 
\frac{1}{4}\left(2 C_{{\tilde g}a} 
+ 2 C_{{\tilde g}c} 
+ C_{{\tilde g}b}\right) {\cal O}^{RR}_{2} \right].
\end{eqnarray}
\underline{$\chi$-{\it exchange contribution}}:
\begin{eqnarray}\label{eq:Neutralino-1}
{\cal L}_{eff}^{\tilde{g}} &=&
\frac{G_F^2}{2 m_{_p}} \sum\limits_{i=a\cdots f} C_{\#i}\ {\cal O}_{\#i}
=\\
\nonumber
&=& \frac{G_F^2}{2 m_{_p}} \frac{1}{128}
\left[4\left(C_{b} +  C_{c} +  C_{a} + 4 C_{f} - 2 C_{d} - 2 C_{e} \right) {\cal O}^{RR}_{1} +\right.\\
\nonumber
&+&\left.\left(C_{b} -  C_{c} - C_{a}\right) {\cal O}^{RR}_{2} \right].
\end{eqnarray}
The Wilson coefficients were calculated in Ref. \cite{Hirsch:1995ek}:
\begin{eqnarray}\label{eq:PRD53-Gluino}
&&C_{{\tilde g}c} = \frac{\kappa_{3} }{m_{\tilde{g}}} 
\frac{1}{m^{4}_{\tilde{u}_{L}}}, \ 
C_{{\tilde g}a} = \frac{\kappa_{3} }{m_{\tilde{g}}} 
\frac{1}{m^{4}_{\tilde{d}_{R}}},\ 
C_{{\tilde g}b} =  -  \frac{\kappa_{3} }{m_{\tilde{g}}} 
\frac{1}{m^{2}_{\tilde{u}_{L}} m^{2}_{\tilde{d}_{R}}}, \\
\label{eq:PRD53-Neutralino-1}
&& C_{b} =  \frac{\kappa_{2}}{m_{\chi}} \frac{\epsilon_{L} (u) \epsilon_{R} (d)}{m^{2}_{\tilde{u}_{L}}m^{2}_{\tilde{d}_{R}}},\ 
C_{c} =  \frac{\kappa_{2}}{m_{\chi}} \frac{\epsilon^{2}_{L} (u) }{m^{4}_{\tilde{u}_{L}}},\
C_{a} =  \frac{\kappa_{2}}{m_{\chi}} \frac{\epsilon^{2}_{R} (d) }{m^{4}_{\tilde{d}_{R}}},\\
\label{eq:PRD53-Neutralino-2} 
&&C_{f} = \frac{\kappa_{2}}{m_{\chi}} \frac{\epsilon^{2}_{L} (e) }{m^{4}_{\tilde{e}_{L}}},\ 
C_{d} =  \frac{\kappa_{2}}{m_{\chi}} \frac{\epsilon_{L} (e) \epsilon_{R} (d)}{m^{2}_{\tilde{e}_{L}}m^{2}_{\tilde{d}_{R}}},\ 
C_{e} =  \frac{\kappa_{2}}{m_{\chi}} \frac{\epsilon_{L} (e) \epsilon_{L} (u)}{m^{2}_{\tilde{e}_{L}}m^{2}_{\tilde{u}_{L}}},
\end{eqnarray}
with 
\begin{eqnarray}\label{eq:Defs-RPV-1}
&& \kappa_{2} = \lambda'^{2}_{111}4\pi \alpha_{2} \frac{m_{p}}{G_{F}^{2}}, \hspace{31mm}   
\kappa_{3} = \lambda'^{2}_{111} 16 \pi \alpha_{s} \frac{m_{p}}{G_{F}^{2}},\\
&& \epsilon_{L}(\psi) 
= \tan\theta_{W} [{T_{3}(\psi) - Q(\psi)}] , \ \ 
\epsilon_{R}(\psi) =\tan\theta_{W} Q(\psi)\, ,
\end{eqnarray}
where $\lambda'_{111}$ is the first generation \rp SUSY coupling,
$\alpha_{2} = g^{2}_{2}/4\pi$ and $\alpha_{s} = g^{2}_{3}/4\pi$ are
the $SU(2)_{L}$ and $SU(3)_{C}$ couplings, respectively.  As usual
$G_{F}$ is the Fermi constant and $m_{p}$ is the proton
mass. $T_{3}(\psi)$ and $Q(\psi)$ are the third component of the weak
isospin and the electric charge of the fermion $\psi$.

First we consider the $\tilde{g}$-exchange and derive the limits on
the \rp SUSY parameter space.  For this we adopt the conventional
assumption $m_{\tilde{u}_{L}}\approx m_{\tilde{d}_{R}} \approx
m_{\tilde{q}}$. Comparing the Lagrangian (\ref{eq:Gluino-1}) with the
canonic form (\ref{eq:LagGen}) and using the half-life formula
(\ref{eq:T12}) we find, by taking into account the QCD running, for
the current experimental limits (\ref{eq:T0nu-1})-(\ref{eq:T0nu-2})
the following upper bounds on the \rp SUSY Yukawa coupling:
\begin{eqnarray}\label{eq:Lambda-Lim-QCD-1}
\tilde{g}-\mbox{exchange}:&&\lambda_{111Ge}^{'}\leq 1.0\times 10^{-2}
\left( \frac{m_{\tilde{q}}}{\mbox{1TeV}} \right)^{2}\left(\frac{m_{\tilde{g}}}{\mbox{1TeV}}\right)^{1/2}, \\
\label{eq:Lambda-Lim-QCD-2}
&& \lambda_{111Xe}^{'}\leq 7.2\times 10^{-3}\left( \frac{m_{\tilde{q}}}{\mbox{1TeV}} \right)^{2}\left(\frac{m_{\tilde{g}}}{\mbox{1TeV}}\right)^{1/2}
\end{eqnarray}

For the case of the neutralino exchange we consider a particular part
of the \rp SUSY parameter space where $m_{\tilde{e}}\ll
m_{\tilde{q}}$. This is motivated by the fact that LHC searches set
very strong limits on the colored sector of any beyond the SM physics.
In this domain the dominant contribution comes from the diagram (f),
corresponding to eHSM $\#23$. We find the limits taking into account
the QCD running

\begin{eqnarray}\label{eq:Lambda-Lim-QCD-3}
\chi-\mbox{exchange}:&&\lambda_{111Ge}^{'}\leq 7.3\times 10^{-1}
\left( \frac{m_{\tilde{e}}}{\mbox{1TeV}} \right)^{2}
\left(\frac{m_{\tilde{\chi}}}{\mbox{1TeV}}\right)^{1/2}\\
\label{eq:Lambda-Lim-QCD-4}
&&\lambda_{111 Xe}^{'}\leq 5.1
\times 10^{-1}\left( \frac{m_{\tilde{e}}}{\mbox{1TeV}} \right)^{2}
\left(\frac{m_{\tilde{\chi}}}{\mbox{1TeV}}\right)^{1/2} 
\end{eqnarray}
For the calculation of this limit we assumed $N_1 \simeq 1$.  Note
that the limits from the $\chi$-exchange (\ref{eq:Lambda-Lim-QCD-3}),
(\ref{eq:Lambda-Lim-QCD-4}) are competitive with those, which come
from the $\tilde{g}$-exchange (\ref{eq:Lambda-Lim-QCD-1}),
(\ref{eq:Lambda-Lim-QCD-2}) in the \rp SUSY parameter space domain
$m_{\tilde{q}}\gg m_{\tilde{e}}$ and $m_{\tilde{g}}\gg m_{\chi}$.

In order to demonstrate the significance of the QCD running we
re-calculated the corresponding limit for ${}^{76}$Ge using the same
experimental bound (\ref{eq:T0nu-1}), but switching off the QCD
corrections.  This results in a modification of the coefficients
$\beta$ in Table~\ref{t:betaparameters}, which can be found for this
limiting case in Ref.~\cite{Gonzalez:2015ady}.  Without QCD running we
obtain the limits:
\begin{eqnarray}\label{eq:GluinoNoQCD}
\tilde{g}-\mbox{exchange}:&&\lambda_{111Ge}^{'}\leq 9.3\times 10^{-2}\left( \frac{m_{\tilde{q}}}{\mbox{1TeV}} \right)^{2}\left(\frac{m_{\tilde{g}}}{\mbox{1TeV}}\right)^{1/2},\\
\label{eq:CharginoNoQCD}
\chi-\mbox{exchange}:&& \lambda_{111Ge}^{'}\leq 5.2\ \  \left( \frac{m_{\tilde{e}}}{\mbox{1TeV}} \right)^{2}\left(\frac{m_{\tilde{\chi}}}{\mbox{1TeV}}\right)^{1/2} 
\end{eqnarray}
This is about $\sim$10 ($\sim$7) weaker than the limits for gluino
(neutralino) cases in Eqs.~(\ref{eq:Lambda-Lim-QCD-1}),
(\ref{eq:Lambda-Lim-QCD-3}) taking into account the QCD running. This
again demonstrates the crucial role of the QCD corrections for SRM.

\section{Conclusions}
\label{sec:Discussion}

In this paper we have calculated QCD-improved lower limits on the
Wilson coefficients and the LNV mass scales, $\Lambda_{LNV}$, for all
ultraviolet completions (``elementary high-scale models'') of the
$d=9$ $\znbb$ decay operator, contributing to the short-range part of
the amplitude.  We have also worked out a general method which can be
used to find the limits for any particular model, contributing to the
SRM of $\znbb$ decay with several diagrams. Our method can be used to
find new, improved limits easily, should better experimental limits or
new calculations of the nuclear matrix elements become available.  In
closing, we would like to stress again, that QCD running can lead to
important changes in the $\znbb$ decay limits on the mass scales of
LNV extensions of the SM.

\bigskip

\centerline{\bf Acknowledgements}

\medskip

Marcela and Carolina are grateful for the hospitality of the AHEP
group in the IFIC during their visits in May-July 2016. This work was
supported by the Spanish MICINN grants SEV-2014-0398, FPA2014-58183-P
and Multidark CSD2009-00064 (MINECO), and PROMETEOII/2014/084
(Generalitat Valenciana), and by Fondecyt (Chile) under grants
No. 3150472, No. 1150792 and No. 3160642 as well as CONICYT (Chile)
Ring ACT 1406 and Basal FB0821. M.H. thanks the Universidad Federico
Santamaria, Valparaiso, for hospitality.

%%%%%%%%%%%%%%%%%%%%%%%%%%%%%%%%%%%%%%%%%%%%%%%%%%%%%%%%%%%%%%%%%%%%%%

\appendix

\section{Specification of eHSMs and Notations}
\label{sec:App:Decomp}

Here we comment on the notations used in
Tables~\ref{t:HSMIdentTI-1}-\ref{t:HSMIdentTII} where we specify all
the eHSMs contributing to $\znbb$-decay via the short-range mechanism
according to T-I and T-II diagrams in Fig.~\ref{fig:short} with heavy
intermediate particles (messengers).  Each eHSM is uniquely specified
by the SM gauge group $G_{\rm SM} = SU(3)_{c}\times SU(2)_{L}\times
U(1)_{Y}$ assignments of the messengers: Scalar-Fermion-Scalar $\{(S),
(\psi), (S')\}$ for diagram T-I and triple scalar $\{(S), (S'),
(S'')\}$ for T-II. Thus each set of $G_{\rm SM}$ representations in
curled brackets corresponds to a particular eHSM = $\{(), (), ()\}$.
The list of the models is taken from Ref.~\cite{Bonnet:2012kh},
however, in our tables we put the eHSMs in groups with an identifier
\#I. The eHSMs from the same group lead in the low-energy limit after
integrating out the heavy particles to the same effective operator
${\cal O}^{I}$.  These operators in the form
(\ref{eq:OI-basis-decomp-1}) are given in
Tables~\ref{t:HSModelsShort-range-1}-\ref{t:HSModelsShort-range-3}.

Some eHSMs appear in Tables~\ref{t:HSMIdentTI-1}-\ref{t:HSMIdentTII}
with numerical coefficients $\alpha$, like $\alpha\cdot \{(),(),()\}$.
In our notations this means that in the low-energy limit the models
belonging to the group \#I tend to the same effective operator ${\cal
  O}^{I}$ but with different normalization factors so that
\begin{eqnarray}\label{eq:App-5}
\left.\begin{array}{r}
\alpha \cdot \{(),(),()\}   \\
\{(),(),()\}
\end{array}
\right\} \rightarrow \mathcal{O}^{I}
\end{eqnarray}
For example, in the group \#1 we find the eHSMs for which 
\begin{eqnarray}\label{eq:App-1}
-2\cdot \{(),(),()\}&:&     \{({\bf 8}, {\bf 2}; 1/2),({\overline{\bf 3}}, {\bf 2}; 5/6),({\overline{\bf 3}}, 
{\bf 1}; 1/3)\} \rightarrow -\frac{1}{2}\cdot \mathcal{O}^{I=1}\\
\label{eq:App-2}
\{(),(),()\}&:& 
\{({\bf 8}, {\bf 2}; 1/2),({\bf 8}, {\bf 1}; 0),({\bf 8}, {\bf 2}; -1/2)\}
\hspace{1mm}  \rightarrow \hspace{8mm}\hspace{1mm} \mathcal{O}^{I=1}
\end{eqnarray}
We also used a shorthand notation for the subsets of eHSMs in a
particular group \#I inside the blue boldface curled brackets, which
means
\begin{eqnarray}\label{eq:App3}
&& \alpha\cdot \textbf{{\color{blue}\{}}\{(), (), ()\},..., \{(), (), ()\}... \textbf{{\color{blue}
\}}} = 
\alpha\cdot \{(), (), ()\},..., \alpha\cdot\{(), (), ()\}...
\end{eqnarray}
Limits on the Wilson coefficients $C_{I}$ of the eHSMs and their
characteristic mass scales $M_{I}$ are given in
Tables~\ref{t:HSModelsShort-range-1}-\ref{t:HSModelsShort-range-3} for
each eHSM listed in Tables~\ref{t:HSMIdentTI-1}-\ref{t:HSMIdentTII}.
For an eHSM appearing in the latter Tables with a numerical coefficient 
eHSM = $\alpha\cdot\{(), (), ()\}$ the upper limit from
Tables~\ref{t:HSModelsShort-range-1}-\ref{t:HSModelsShort-range-3} on
its Wilson coefficient should be replaced with $\alpha\cdot
C_{I}^{exp}$ and the lower limit on the mass scale with $\alpha^{-1/5}
\overline{M}_{I}^{exp}$.

We refer the reader to Ref.~\cite{Bonnet:2012kh} for the detailed
rules of the reconstruction of the operators ${\cal O}^{I}$ starting
from the eHSM messenger assignment $\{(),(),()\}$ given in
Tables~\ref{t:HSMIdentTI-1}-\ref{t:HSMIdentTII}.

\hspace{-0.5 cm}
\begin{table}[h]
%{\footnotesize 

\resizebox{\textwidth}{8cm}{  
\begin{tabular}{c|c}
%\scalebox{0.3}
\hline\hline
&T-I
%\multicolumn{4}{l||}{Set eHSM  \hspace{30mm} $\mathcal{O}^{I } = x_{i}^{XY} \ \%mathcal{O}_{i}^{XY}+y_{j}^{XY} \ \mathcal{O}_2^{XX}$} 
\\
\hline
%\cline{3-6}
eHSM&\hspace{18mm}  Mediators $(SU(3)_{c}, SU(2)_{L}, U(1)_{Y})$ \ \ \ \ \ with $Y = Q - T_{3}$
\\
%:{SK_158} A modification
%\#I & $\{(S, \Psi, S^{'})\}$\\
\#I & $\{(S), (\psi), (S^{'})\}$\\
%:{SK_158}  END
\hline
%
% 
%
%&$x \ \mathcal{O}_1^{XX}+y \ \mathcal{O}_2^{XX}$&&&&&&\\
1&$\{({\bf 8}, {\bf 2}; 1/2) , ({\bf 8}, {\bf 1}; 0), ({\bf 8}, {\bf 2}; -1/2)\}, \{({\bf 8}, {\bf 2}; 1/2) ,({\bf 8}, {\bf 3}; 0),({\bf 8}, {\bf 2}; -1/2)\}, \{({\bf 8}, {\bf 2}; 1/2),({\bf 3}, {\bf 3}; 2/3), ({\bf 1}, {\bf 3}; 1)\},      $ \\

 & $ \{({\bf 8}, {\bf 2}; 1/2),({\bf 3}, {\bf 2}; 7/6), ({\bf 1}, {\bf 3}; 1)\},    \{({\bf 8}, {\bf 2}; 1/2), ({\overline{\bf 3}}, {\bf 2}; 5/6), ({\bf 1}, {\bf 3}; 1)\},  \{({\bf 8}, {\bf 2}; 1/2), ({\overline{\bf 3}}, {\bf 3}; 1/3), ({\bf 1}, {\bf 3}; 1)\}$\\

& $-2\cdot \textbf{{\color{blue}\{}}\{({\bf 8}, {\bf 2}; 1/2),({\overline{\bf 3}}, {\bf 2}; 5/6),({\overline{\bf 3}}, {\bf 1}; 1/3)\} , \{({\bf 8}, {\bf 2}; 1/2),({\overline{\bf 3}}, {\bf 2}; 5/6),({\overline{\bf 3}}, {\bf 3}; 1/3)\},       \{({\bf 8}, {\bf 2}; 1/2),({\bf 8}, {\bf 1}; 0),({\overline{\bf 3}}, {\bf 1}; 1/3)\}, $ \\

& $ \{({\bf 8}, {\bf 2}; 1/2), ({\bf 8}, {\bf 3}; 0), ({\overline{\bf 3}}, {\bf 3}; 1/3)\}        , \{({\bf 8}, {\bf 2}; 1/2), ({\bf 3}, {\bf 3}; 2/3), ({\bf 3}, {\bf 2}; 1/6)\}, 
\{({\bf 8}, {\bf 2}; 1/2),({\bf 8}, {\bf 1}; 0),({\bf 3}, {\bf 2}; 1/6) \}$\\

& $\{({\bf 8}, {\bf 2}; 1/2),({\bf 8}, {\bf 3}; 0),({\bf 3}, {\bf 2}; 1/6)\} \textbf{{\color{blue}\}}}$ \\

2& $\{({\overline{\bf 3}}, {\bf 2}; -1/6),({\bf 1}, {\bf 1}; 0),({\overline{\bf 3}}, {\bf 1}; 1/3)\}, \{({\overline{\bf 3}}, {\bf 2}; -1/6),({\bf 1}, {\bf 3}; 0),({\overline{\bf 3}}, {\bf 3}; 1/3)\}$ \\

3& $\{({\overline{\bf 3}}, {\bf 2}; -1/6),({\bf 8}, {\bf 1}; 0),({\overline{\bf 3}}, {\bf 1}; 1/3)\}, \{({\overline{\bf 3}}, {\bf 2}; -1/6),({\bf 8}, {\bf 3}; 0),({\overline{\bf 3}}, {\bf 3}; 1/3)\} $\\

4& $\{({\overline{\bf 3}}, {\bf 2}; -1/6),({\bf 1}, {\bf 1}; 0),({\bf 3}, {\bf 2}; 1/6)\}, \{({\overline{\bf 3}}, {\bf 2}; -1/6),({\bf 1}, {\bf 3}; 0),({\bf 3}, {\bf 2}; 1/6)\}, \{({\bf 6}, {\bf 3}; 1/3) ,({\bf 3}, {\bf 3}; 2/3) ,({\bf 3}, {\bf 2}; 1/6)\},$\\

& $\{({\bf 6}, {\bf 3}; 1/3) ,({\bf 6}, {\bf 2}; -1/6),({\bf 3}, {\bf 2}; 1/6)\}, \{({\bf 3}, {\bf 1}; -1/3),({\bf 1}, {\bf 1}; 0) ,({\overline{\bf 3}}, {\bf 1}; 1/3)\},  \{({\bf 3}, {\bf 3}; -1/3),({\bf 1}, {\bf 3}; 0) ,({\overline{\bf 3}}, {\bf 3}; 1/3)\} $\\

& $  \{({\bf 3}, {\bf 1}; -1/3),({\bf 6}, {\bf 2}; -1/6) ,({\bf 6}, {\bf 1}; -2/3)\}     ,    \{({\bf 3}, {\bf 3}; -1/3),({\bf 6}, {\bf 2}; -1/6) ,({\bf 6}, {\bf 1}; -2/3)\},   \{({\bf 3}, {\bf 1}; -1/3) ,({\bf 3}, {\bf 2}; -5/6),({\bf 6}, {\bf 1}; -2/3)\}$  \\

& $  \{({\bf 3}, {\bf 3}; -1/3),({\bf 3}, {\bf 2}; -5/6) ,({\bf 6}, {\bf 1}; -2/3)\} $\\

& $-2 \cdot \textbf{{\color{blue}\{}}\{({\overline{\bf 3}}, {\bf 2}; -1/6) ,({\overline{\bf 6}}, {\bf 2}; 1/6),({\overline{\bf 3}}, {\bf 1}; 1/3)\}, \{({\overline{\bf 3}}, {\bf 2}; -1/6) ,({\overline{\bf 6}}, {\bf 2}; 1/6),({\overline{\bf 3}}, {\bf 3}; 1/3)\}\textbf{{\color{blue}\}}}$  \\

%%%%%%%%%%%%%%%%%%%
& $-\frac{1}{2}\cdot \textbf{{\color{blue}\{}}\{({\bf 6}, {\bf 3}; 1/3) ,({\bf 6}, {\bf 2}; -1/6),({\bf 6}, {\bf 1}; -2/3)\}, \{({\bf 6}, {\bf 1}; 4/3) ,({\bf 6}, {\bf 2}; 5/6),({\bf 6}, {\bf 3}; 1/3)\}, \{({\bf 6}, {\bf 3}; 1/3) ,({\bf 3}, {\bf 3}; 2/3),({\bf 1}, {\bf 3}; 1),\}$  \\

& $\{({\bf 6}, {\bf 1}; 4/3) ,({\bf 3}, {\bf 2}; 7/6),({\bf 1}, {\bf 3}; 1)\}, \{({\overline{\bf 6}}, {\bf 3}; -1/3) ,({\overline{\bf 3}}, {\bf 3}; 1/3),({\bf 1}, {\bf 3}; 1)\}, \{({\overline{\bf 6}}, {\bf 1}; 2/3) ,({\overline{\bf 3}}, {\bf 2}; 5/6),({\bf 1}, {\bf 3}; 1)\}\textbf{{\color{blue}\}}}$  \\

5& $\{({\overline{\bf 3}}, {\bf 2}; -1/6), ({\bf 8}, {\bf 1}; 0),({\bf 3}, {\bf 2}; 1/6)\},  \{({\overline{\bf 3}}, {\bf 2}; -1/6), ({\bf 8}, {\bf 3}; 0),({\bf 3}, {\bf 2}; 1/6)\} ,\{({\bf 3}, {\bf 1}; -1/3),({\bf 8}, {\bf 1}; 0),({\overline{\bf 3}}, {\bf 1}; 1/3)\}$ \\

& $\{({\bf 3}, {\bf 3}; -1/3), ({\bf 1}, {\bf 3}; 0),({\overline{\bf 3}}, {\bf 3}; 1/3)\}$ \\

6&$\{(\overline{{\bf 3}}, {\bf 2}; -1/6) , ({\bf 3}, {\bf 2}; 1/6) , (\overline{{\bf 3}}, {\bf 1}; 1/3)\}, \{(\overline{{\bf 3}}, {\bf 2}; -1/6) , ({\bf 3}, {\bf 2}; 1/6) , (\overline{{\bf 3}}, {\bf 3}; 1/3)\}$ \\

\hline 
7&$\{({\bf 8}, {\bf 2}; 1/2) , ({\bf 8}, {\bf 1}; 0), ({\bf 8}, {\bf 2}; -1/2)\},\{({\bf 8}, {\bf 2}; 1/2),({\bf 8}, {\bf 3}; 0),({\bf 8}, {\bf 2}; -1/2)\}, \{({\bf 8}, {\bf 2}; 1/2) , ({\bf 3}, {\bf 2}; 7/6) ,({\bf 1}, {\bf 3}; 1)\},$ \\

&$\{({\bf 8}, {\bf 2}; 1/2), ({\bf 3}, {\bf 3}; 2/3), ({\bf 1}, {\bf 3}; 1)\}, \{({\bf 8}, {\bf 2}; 1/2) , ({\overline{\bf 3}}, {\bf 3}; 1/3), ({\bf 1}, {\bf 3}; 1)\},\{({\bf 8}, {\bf 2}; 1/2) , ({\overline{\bf 3}}, {\bf 2}; 5/6) , ({\bf 1}, {\bf 3}; 1)\}$ \\

&$-2 \cdot \textbf{{\color{blue}\{}}\{({\bf 8}, {\bf 2}; 1/2),({\overline{\bf 3}}, {\bf 2}; 5/6),({\overline{\bf 3}}, {\bf 1}; 1/3)\}, 
\{({\bf 8}, {\bf 2}; 1/2), ({\overline{\bf 3}}, {\bf 2}; 5/6) ,({\overline{\bf 3}}, {\bf 3}; 1/3)\},\{({\bf 8}, {\bf 2}; 1/2), ({\bf 8}, {\bf 1}; 0)({\overline{\bf 3}}, {\bf 1}; 1/3) \},$ \\

&$\{ ({\bf 8}, {\bf 2}; 1/2)  ,({\bf 8}, {\bf 3}; 0), ({\overline{\bf 3}}, {\bf 3}; 1/3)\},\{({\bf 8}, {\bf 2}; 1/2), ({\bf 3}, {\bf 3}; 2/3),({\bf 3}, {\bf 2}; 1/6)\},\{({\bf 8}, {\bf 2}; 1/2), ({\bf 8}, {\bf 1}; 0),({\bf 3}, {\bf 2}; 1/6)\},$ \\

&$\{({\bf 8}, {\bf 2}; 1/2),({\bf 8}, {\bf 3}; 0),({\bf 3}, {\bf 2}; 1/6)\}\textbf{{\color{blue}\}}}$ \\
\\ 
\hline
\hline

\end{tabular}
}
\caption{ Identification of the T-I (Fig.~\ref{fig:short}) short-range
  eHSMs. For explanation of notations see
  Appendix~\ref{sec:App:Decomp} and the main text.}
\label{t:HSMIdentTI-1}   
\end{table}
%
%%%%%%%%%%%%%%%%% TI-2 %%%%%%%%%
%
\hspace{-0.5 cm}
\begin{table}[h]
%{\footnotesize 

\resizebox{\textwidth}{7cm}{  
\begin{tabular}{c|c}
%\scalebox{0.3}
\hline\hline
& T-I
%\multicolumn{4}{l||}{Set eHSM  \hspace{30mm} $\mathcal{O}^{I } = x_{i}^{XY} \ \%mathcal{O}_{i}^{XY}+y_{j}^{XY} \ \mathcal{O}_2^{XX}$} 
\\
\hline
%\cline{3-6}
eHSM& \hspace{18mm}  Mediators $(SU(3)_{c}, SU(2)_{L}, U(1)_{Y})$ \ \ \ \ \ with $Y = Q - T_{3}$
\\
%:{SK_155} A modification
%\#I & $\{(S, \Psi, S^{'})\}$\\
\#I & $\{(S), (\psi), S^{'})\}$\\
%:{SK_155}  END
\hline
%
% 
%
%%%%%%%%%%%%%%%%%%%%%%%%%
%
8& $\{(\overline{{\bf 3}}, {\bf 2}; -1/6),({\bf 8}, {\bf 2}; 1/2),(\overline{{\bf 3}}, {\bf 1}; 1/3)\},  \{(\overline{{\bf 3}}, {\bf 2}; -1/6),({\bf 8}, {\bf 2}; 1/2),(\overline{{\bf 3}}, {\bf 3}; 1/3)\     $ 
\\
9&   $\{(\overline{{\bf 3}}, {\bf 2}; -1/6),({\bf 3}, {\bf 1}; -1/3),(\overline{{\bf 3}}, {\bf 1}; 1/3)\}, \{({\overline{\bf 3}}, {\bf 2}; -1/6),({\bf 3}, {\bf 3}; -1/3),(\overline{{\bf 3}}, {\bf 3}; 1/3)\}$  \\
10&  $\{(\overline{{\bf 3}}, {\bf 2}; -1/6),({\overline{\bf 6}}, {\bf 1}; -1/3),(\overline{{\bf 3}}, {\bf 1}; 1/3)\}, \{(\overline{{\bf 3}}, {\bf 2}; -1/6),({\overline{\bf 6}}, {\bf 3}; -1/3),(\overline{{\bf 3}}, {\bf 3}; 1/3)\} $       \\
\hline
11&  $\{({\bf 8}, {\bf 2}; 1/2),(\overline{{\bf 3}}, {\bf 2}; 5/6) ,(\overline{{\bf 3}}, {\bf 1};1/3)\},\{({\bf 8}, {\bf 2}; 1/2) , ({\bf 8}, {\bf 1}; 0) ,(\overline{{\bf 3}}, {\bf 1};1/3)\},\{({\bf 8}, {\bf 2}; 1/2), ({\bf 3}, {\bf 2}; 7/6) , ({\bf 3}, {\bf 2};1/6)\}$  \\
&  $\{({\bf 8}, {\bf 2}; 1/2) , ({\bf 8}, {\bf 2}; -1/2) , ({\bf 3}, {\bf 2};1/6)\}$  \\
12& $\{(\overline{{\bf 3}}, {\bf 2}; -1/6) , ({\bf 1}, {\bf 1}; 0) , (\overline{{\bf 3}}, {\bf 1}; 1/3)\}$        
\\
13& $\{({\bf 3}, {\bf 1}; -1/3) , ({\bf 1}, {\bf 1}; 0) , (\overline{{\bf 3}}, {\bf 1}; 1/3)\},\{({\bf 3}, {\bf 1}; -1/3) , ({\bf 6}, {\bf 1}; 1/3) ,({\bf 6}, {\bf 1}; -2/3)\},\{({\bf 3}, {\bf 1}; -1/3), ({\bf 6}, {\bf 2}; -1/6) , ({\bf 6}, {\bf 1}; -2/3)\}$            \\
& $\{({\bf 3}, {\bf 1}; -1/3),({\bf 3}, {\bf 1}; -4/3) ,({\bf 6}, {\bf 1}; -2/3)\},\{({\bf 3}, {\bf 1}; -1/3), ({\bf 3}, {\bf 2}; -5/6) , ({\bf 6}, {\bf 1}; -2/3)\}$ \\
& $-2 \cdot \textbf{{\color{blue}\{}}\{(\overline{{\bf 3}}, {\bf 2}; -1/6), ({\overline{\bf 6}}, {\bf 2}; 1/6), (\overline{{\bf 3}}, {\bf 1}; 1/3)\}\textbf{{\color{blue}\}}}$ \\  
14& $\{(\overline{{\bf 3}}, {\bf 2}; -1/6), ({\bf 8}, {\bf 1}; 0), (\overline{{\bf 3}}, {\bf 1}; 1/3)\}$   
\\
15&  $\{(\overline{{\bf 3}}, {\bf 2}; -1/6) , ({\bf 3}, {\bf 2}; 1/6) , (\overline{{\bf 3}}, {\bf 1}; 1/3)\}$  
\\
16& $\{({\bf 3}, {\bf 1}; -1/3), ({\bf 8}, {\bf 1}; 0),(\overline{{\bf 3}}, {\bf 1}; 1/3)\}$   \\
\hline
17& $\{({\bf 8}, {\bf 2}; 1/2) ,({\overline{\bf 3}}, {\bf 2}; 5/6),({\overline{\bf 3}}, {\bf 1}; 1/3)\} ,\{({\bf 8}, {\bf 2}; 1/2), ({\bf 8}, {\bf 1}; 0) ,(\overline{{\bf 3}}, {\bf 1}; 1/3)\},\{({\bf 8}, {\bf 2}; 1/2) , ({\bf 3}, {\bf 2}; 7/6) , ({\bf 3}, {\bf 2}; 1/6)\}$      \\
& $\{({\bf 8}, {\bf 2}; 1/2) , ({\bf 8}, {\bf 2}; -1/2) , ({\bf 3}, {\bf 2}; 1/6)\}$      \\
18&  $\{({\overline{\bf 3}}, {\bf 2}; -1/6) , ({\bf 1}, {\bf 2}; 1/2),({\overline{\bf 3}}, {\bf 1}; 1/3)\}$     
\\
19& $\{(\overline{{\bf 3}}, {\bf 2}; -1/6) , ({\bf 8}, {\bf 2}; 1/2) , (\overline{{\bf 3}}, {\bf 1}; 1/3)\}$       
\\

& $\{(\overline{{\bf 3}}, {\bf 2}; -1/6), ({\bf 8}, {\bf 1}; 0), ({\overline{\bf 3}}, {\bf 1}; 1/3)\}$       \\
20&  $\{(\overline{{\bf 3}}, {\bf 2}; -1/6), ({\bf 3}, {\bf 1}; -1/3),(\overline{{\bf 3}}, {\bf 1}; 1/3)\}$    
\\
%
%%%%%%%%%%%%%%%%%%%%%%%%%
\\ 
\hline
\hline

\end{tabular}
}
\caption{Continuation of Table~\ref{t:HSMIdentTI-1}.}
\label{t:HSMIdentTI-2}
\end{table}

%%%%%
\hspace{-0.5 cm}
\begin{table}[h]
%{\footnotesize 

\resizebox{\textwidth}{8cm}{  
\begin{tabular}{c|c}
%\scalebox{0.3}
\hline\hline
& T-I
%\multicolumn{4}{l||}{Set eHSM  \hspace{30mm} $\mathcal{O}^{I } = x_{i}^{XY} \ \%mathcal{O}_{i}^{XY}+y_{j}^{XY} \ \mathcal{O}_2^{XX}$} 
\\
\hline
%\cline{3-6}
eHSM& \hspace{18mm}  Mediators $(SU(3)_{c}, SU(2)_{L}, U(1)_{Y})$ \ \ \ \ \ with $Y = Q - T_{3}$
\\
%:{SK_156} A modification
%\#I & $\{(S, \Psi, S^{'})\}$\\
\#I & $\{(S), (\psi), (S^{'})\}$\\
%:{SK_156}  END
\hline
21& $  \{({\overline{\bf 3}}, {\bf 2}; -7/6), ({\bf 1}, {\bf 2}; -1/2) ,({\bf 3}, {\bf 2}; 1/6)\}   ,\{({\bf 6}, {\bf 1}; 4/3) , ({\bf 3}, {\bf 2}; 7/6) , ({\bf 3}, {\bf 2}; 1/6)\}, \{({\bf 6}, {\bf 1}; 4/3) ,({\bf 3}, {\bf 1}; 5/3) , ({\bf 3}, {\bf 2}; 7/6)\} $
\\
& $ \{({\bf 6}, {\bf 1}; 4/3) , ({\bf 6}, {\bf 1}; 1/3), ({\bf 3}, {\bf 2}; 1/6)\}, \{({\bf 6}, {\bf 1}; 4/3) , ({\bf 6}, {\bf 2}; 5/6) , ({\bf 3}, {\bf 2}; 7/6)\} ,   $       \\

& $-2 \cdot \textbf{{\color{blue}\{}}\{(\overline{{\bf 3}}, {\bf 2}; -1/6), ({\overline{\bf 6}}, {\bf 1}; -1/3), (\overline{{\bf 3}}, {\bf 1}; 1/3)\}\textbf{{\color{blue}\}}}$ \\

22& $\{(\overline{{\bf 3}}, {\bf 2}; -7/6), ({\bf 8}, {\bf 2}; -1/2),({\bf 3}, {\bf 2}; 1/6)\}$   \\  

%23&  \\  
\hline
23& $\{({\bf 1}, {\bf 2}; 1/2), ({\bf 1}, {\bf 1}; 0), ({\bf 1}, {\bf 2}; 1/2)\},\{ ({\bf 1}, {\bf 2}; 1/2), ({\bf 1}, {\bf 3}; 0), ({\bf 1}, {\bf 2}; -1/2) \},\{({\bf 1}, {\bf 2}; 1/2), ({\bf 3}, {\bf 3}; 2/3), ({\overline{\bf 1}}, {\bf 3}; 1)\}, $ \\

& $  \{ ({\bf 1}, {\bf 2}; 1/2), ({\overline{\bf 3}}, {\bf 2}; 7/6), ({\bf 1}, {\bf 3}; 1)\}   ,\{ ({\bf 1}, {\bf 2}; 1/2), ({\overline{\bf 3}}, {\bf 2}; 5/6), ({\bf 1}, {\bf 3}; 1)\}, \{ ({\bf 1}, {\bf 2}; 1/2), ({\overline{\bf 3}}, {\bf 3}; 1/3), ({\bf 1}, {\bf 3}; 1)\}$\\

%& $-\frac{1}{2}\{\{({\bf 1}, {\bf 2}; 1/2), ({\bf 1}, {\bf 1}; 0), (\overline{{\bf 3}}, {\bf 1}; 1/3)\}, \{ ({\bf 1}, {\bf 2}; 1/2), ({\overline{\bf 3}}, {\bf 2}; 5/6), ({\overline{\bf 3}}, {\bf 3}; 1/3)\}, \{({\bf 1}, {\bf 2}; 1/2), ({\overline{\bf 3}}, {\bf 2}; 5/6), ({\overline{\bf 3}}, {\bf 1}; 1/3), \}$ \\

%& $\{({\bf 1}, {\bf 2}; 1/2), ({\bf 3}, {\bf 3}; 2/3), ({\bf 3}, {\bf 2}; 1/6)\} \}$  \\

& $ -2 \cdot \textbf{{\color{blue}\{}}  \{ ({\bf 1}, {\bf 2}; 1/2), ({\overline{\bf 3}}, {\bf 2}; 5/6), ({\overline{\bf 3}}, {\bf 1}; 1/3)\}, \{({\bf 1}, {\bf 2}; 1/2), ({\overline{\bf 3}}, {\bf 2}; 5/6),({\overline{\bf 3}}, {\bf 3}; 1/3)\}, \{({\bf 1}, {\bf 2}; 1/2), ({\bf 1}, {\bf 1}; 0), ({\overline{\bf 3}}, {\bf 1}; 1/3)\},$\\

&   $\{({\bf 1}, {\bf 2}; 1/2), ({\bf 1}, {\bf 3}; 0), ({\overline{\bf 3}}, {\bf 3}; 1/3)\}, \{({\bf 1}, {\bf 2}; 1/2),({\bf 3}, {\bf 3}; 2/3), ({\bf 3}, {\bf 2}; 1/6)\}, \{({\bf 1}, {\bf 2}; 1/2), ({\bf 1}, {\bf 1}; 0),({\bf 3}, {\bf 2}; 1/6)\}, $  \\

&   $\{({\bf 1}, {\bf 2}; 1/2), ({\bf 1}, {\bf 3}; 0), ({\bf 3}, {\bf 2}; 1/6)\} \textbf{{\color{blue}\}}}$ \\
\hline
24&  $\{ ({\bf 1}, {\bf 2}; 1/2),  ({\bf 1}, {\bf 1}; 0), ({\bf 1}, {\bf 2}; -1/2)\}, \{ ({\bf 1}, {\bf 2}; 1/2),  ({\bf 1}, {\bf 3}; 0), ({\bf 1}, {\bf 2}; -1/2)\}, \{({\bf 1}, {\bf 2}; 1/2), ({\bf 3}, {\bf 2}; 7/6),  ({\bf 1}, {\bf 3}; 1)\}, $ \\

& $\{({\bf 1}, {\bf 2}; 1/2), ({\bf 3}, {\bf 3}; 2/3), ({\bf 1}, {\bf 3}; 1)\}, \{ ({\bf 1}, {\bf 2}; 1/2), ({\overline{\bf 3}}, {\bf 3}; 1/3), ({\bf 1}, {\bf 3}; 1)\}, \{({\bf 1}, {\bf 2}; 1/2), ({\overline{\bf 3}}, {\bf 2}; 5/6), ({\bf 1}, {\bf 3}; 1)\}$  \\

& $ -2 \cdot  \textbf{{\color{blue}\{}}  \{ ({\bf 1}, {\bf 2}; 1/2), ({\overline{\bf 3}}, {\bf 2}; 5/6), ({\overline{\bf 3}}, {\bf 1}; 1/3)\}, \{({\bf 1}, {\bf 2}; 1/2), ({\overline{\bf 3}}, {\bf 2}; 5/6),({\overline{\bf 3}}, {\bf 3}; 1/3)\}, \{({\bf 1}, {\bf 2}; 1/2), ({\bf 1}, {\bf 1}; 0), ({\overline{\bf 3}}, {\bf 1}; 1/3)\},$\\

&   $\{({\bf 1}, {\bf 2}; 1/2), ({\bf 1}, {\bf 3}; 0), ({\overline{\bf 3}}, {\bf 3}; 1/3)\}, \{({\bf 1}, {\bf 2}; 1/2),({\bf 3}, {\bf 3}; 2/3), ({\bf 3}, {\bf 2}; 1/6)\}, \{({\bf 1}, {\bf 2}; 1/2), ({\bf 1}, {\bf 1}; 0),({\bf 3}, {\bf 2}; 1/6)\}, $  \\

&   $\{({\bf 1}, {\bf 2}; 1/2), ({\bf 1}, {\bf 3}; 0), ({\bf 3}, {\bf 2}; 1/6)\} \textbf{{\color{blue}\}}}$ \\
\hline
25& $\{({\overline{\bf 3}}, {\bf 2}; -1/6), ({\bf 1}, {\bf 2}; 1/2), ({\overline{\bf 3}}, {\bf 1}; 1/3)\}, \{({\overline{\bf 3}}, {\bf 2}; -1/6), ({\bf 1}, {\bf 2}; 1/2), ({\overline{\bf 3}}, {\bf 3}; 1/3)\} $ \\

%28& \\
\hline
26&     $\{({\bf 1}, {\bf 2}; 1/2), ({\overline{\bf 3}}, {\bf 2}; 5/6), ({\overline{\bf 3}}, {\bf 1}; 1/3)\}, \{({\bf 1}, {\bf 2}; 1/2),({\bf 1}, {\bf 1}; 0), ({\overline{\bf 3}}, {\bf 1}; 1/3)\}, \{({\bf 1}, {\bf 2}; 1/2), ({\bf 3}, {\bf 2}; 7/6),({\bf 3}, {\bf 2}; 1/6)\}, $ \\

&   $\{({\bf 1}, {\bf 2}; 1/2), ({\bf 1}, {\bf 2}; -1/2), ({\overline{\bf 3}}, {\bf 2}; 1/6)\} \}$ \\
\hline
27&     $\{({\bf 1}, {\bf 2}; 1/2), ({\overline{\bf 3}}, {\bf 2}; 5/6), ({\overline{\bf 3}}, {\bf 1}; 1/3)\}, \{({\bf 1}, {\bf 2}; 1/2),({\bf 1}, {\bf 1}; 0), ({\overline{\bf 3}}, {\bf 1}; 1/3)\}, \{({\bf 1}, {\bf 2}; 1/2), ({\bf 3}, {\bf 2}; 7/6),({\bf 3}, {\bf 2}; 1/6)\}, $ \\

&   $\{({\bf 1}, {\bf 2}; 1/2), ({\bf 1}, {\bf 2}; -1/2), ({\bf 3}, {\bf 2}; 1/6)\} \}$ \\
\hline
28&     $\{({\bf 6}, {\bf 1}; 4/3), ({\bf 6}, {\bf 1}; 1/3), ({\bf 6}, {\bf 1}; -2/3)\}, \{({\bf 6}, {\bf 1}; 4/3),({\bf 3}, {\bf 1}; 5/3), ({\bf 1}, {\bf 1}; 2)\}, \{({\overline{\bf 6}}, {\bf 1}; 2/3), ({\overline{\bf 3}}, {\bf 1}; 4/3),({\bf 1}, {\bf 1}; 2\}, $ \\

 & $-2 \cdot \textbf{{\color{blue}\{}}({\bf 3}, {\bf 1}; -1/3), ({\bf 1}, {\bf 1}; 0), ({\overline{\bf 3}}, {\bf 1}; 1/3)\}, \{({\bf 3}, {\bf 1}; -1/3),({\bf 6}, {\bf 1}; 1/3), ({\bf 6}, {\bf 1}; -2/3)\}, \{({\bf 3}, {\bf 1}; -1/3), ({\bf 3}, {\bf 1}; -4/3),({\bf 6}, {\bf 1}; -2/3\}\textbf{{\color{blue}\}}} $ \\

 & $-\frac{3}{2}\cdot \textbf{{\color{blue}\{}}({\bf 3}, {\bf 1}; -1/3), ({\bf 8}, {\bf 1}; 0), ({\overline{\bf 3}}, {\bf 1}; 1/3)\}\textbf{{\color{blue}\}}} $ \\

\hline
\hline

\end{tabular}
}
\caption{Continuation of Table~\ref{t:HSMIdentTI-1}.}
\label{t:HSMIdentTI-3}
\end{table}

%%%%%%%TOPOLOGY II%%%%%%%%%%%%%%%%%%%%%%%%%%%%%%

\hspace{-0.5 cm}
\begin{table}[h]
%{\footnotesize 

\resizebox{\textwidth}{7cm}{  
\begin{tabular}{c|c}
%\scalebox{0.3}
\hline\hline
& T-II
%\multicolumn{4}{l||}{Set eHSM  \hspace{30mm} $\mathcal{O}^{I } = x_{i}^{XY} \ \%mathcal{O}_{i}^{XY}+y_{j}^{XY} \ \mathcal{O}_2^{XX}$} 
\\
\hline
%\cline{3-6}
eHSM& \hspace{18mm}  Mediators $(SU(3)_{c}, SU(2)_{L}, U(1)_{Y})$ \ \ \ \ \ with $Y = Q - T_{3}$
\\
%:{SK_112}  (S, S^{'}, S^{''})   ?????
\#I & 
%$\{(S, \Psi, S^{'})\}$\\
$\{(S), (S^{'}), (S^{''}) \}$\\
%:{SK_112} END
\hline
%
% 
%
%&$x \ \mathcal{O}_1^{XX}+y \ \mathcal{O}_2^{XX}$&&&&&&\\
1&$\{({\bf 8}, {\bf 2}; 1/2) , ({\bf 8}, {\bf 2}; 1/2), ({\bf 1}, {\bf 3}; -1)\}$ \\

& $-2 \cdot \textbf{{\color{blue}\{}}\{({\bf 8}, {\bf 2}; 1/2) , ({\bf 3}, {\bf 1}; -1/3), ({\overline{\bf 3}}, {\bf 2}; -1/6\}, \{({\bf 8}, {\bf 2}; 1/2) , ({\bf 3}, {\bf 3}; -1/3), ({\overline{\bf 3}}, {\bf 2}; -1/6\}\textbf{{\color{blue}\}}} $\\

5& $-1 \cdot \textbf{{\color{blue}\{}}\{({\bf 6}, {\bf 3}; 1/3) , ({\overline{\bf 6}}, {\bf 1}; 1/3), ({\bf 1}, {\bf 3}; -1)\}, \{({\bf 6}, {\bf 1}; 4/3) , ({\overline{\bf 6}}, {\bf 3}; -1/3), ({\bf 1}, {\bf 3}; -1\}\textbf{{\color{blue}\}}} $\\

 & $2 \cdot \textbf{{\color{blue}\{}}\{({\bf 6}, {\bf 3}; 1/3) , ({\overline{\bf 3}}, {\bf 2}; -1/6), ({\overline{\bf 3}}, {\bf 2}; -1/6)\}, \{({\bf 3}, {\bf 1}; -1/3) , ({\bf 3}, {\bf 1}; -1/3), ({\overline{\bf 6}}, {\bf 1}; 2/3)\},    \{({\bf 3}, {\bf 3}; -1/3) , ({\bf 3}, {\bf 3}; -1/3), ({\overline{\bf 6}}, {\bf 1}; 2/3)\}  \textbf{{\color{blue}\}}}$\\

\hline

7&$\{({\bf 8}, {\bf 2}; 1/2) , ({\bf 8}, {\bf 2}; 1/2), ({\bf 1}, {\bf 3}; -1)\}$ \\

& $-2 \cdot \textbf{{\color{blue}\{}}\{({\bf 8}, {\bf 2}; 1/2) , ({\bf 3}, {\bf 1}; -1/3), ({\overline{\bf 3}}, {\bf 2}; -1/6)\}, \{({\bf 8}, {\bf 2}; 1/2) , ({\bf 3}, {\bf 3}; -1/3), ({\overline{\bf 3}}, {\bf 2}; -1/6)\}  \textbf{{\color{blue}\}}}$\\

\hline

11&$\{({\bf 8}, {\bf 2}; 1/2) , ({\bf 3}, {\bf 1}; 1/3), ({\overline{\bf 3}}, {\bf 2}; -1/6)\}$ \\

16&$2\cdot \textbf{{\color{blue}\{}}({\bf 3}, {\bf 1}; -1/3) , ({\bf 3}, {\bf 1}; -1/3), ({\overline{\bf 6}}, {\bf 1}; 2/3)\textbf{{\color{blue}\}}}$ \\

\hline

17&$\{({\bf 8}, {\bf 2}; 1/2) , ({\bf 3}, {\bf 1}; -1/3), ({\overline{\bf 3}}, {\bf 2}; -1/6)\}$ \\

22&$2 \cdot \textbf {{\color{blue}\{}}({\bf 6}, {\bf 1}; 4/3) , ({\overline{\bf 3}}, {\bf 2}; -7/6), ({\overline{\bf 3}}, {\bf 2}; -1/6)\textbf{{\color{blue}\}}}$ \\

\hline

23&$\{({\bf 1}, {\bf 2}; 1/2) , ({\bf 1}, {\bf 2}; 1/2), ({\bf 1}, {\bf 3}; -1)\}$ \\

& $-2 \cdot \textbf{{\color{blue}\{}}\{({\bf 1}, {\bf 2}; 1/2) , ({\bf 3}, {\bf 1}; -1/3), ({\overline{\bf 3}}, {\bf 2}; -1/6)\}, \{({\bf 1}, {\bf 2}; 1/2) , ({\bf 3}, {\bf 3}; -1/3), ({\overline{\bf 3}}, {\bf 2}; -1/6)\}  \textbf{{\color{blue}\}}}$\\

\hline

24&$\{({\bf 1}, {\bf 2}; 1/2) , ({\bf 3}, {\bf 1}; -1/3), ({\overline{\bf 3}}, {\bf 2}; -1/6)\}, \{({\bf 1}, {\bf 2}; 1/2) , ({\bf 3}, {\bf 3}; -1/3), ({\overline{\bf 3}}, {\bf 2}; -1/6)\}$ \\

&$-\frac{1}{2} \cdot \textbf{{\color{blue}\{}}({\bf 1}, {\bf 2}; 1/2) , ({\bf 1}, {\bf 2}; 1/2), ({\bf 1}, {\bf 3}; -1)\textbf{{\color{blue}\}}}$ \\

\hline

26&$\{({\bf 1}, {\bf 2}; 1/2) , ({\bf 3}, {\bf 1}; -1/3), ({\overline{\bf 3}}, {\bf 2}; -1/6)\}$ \\

\hline

27&$\{({\bf 1}, {\bf 2}; 1/2) , ({\bf 3}, {\bf 1}; -1/3), ({\overline{\bf 3}}, {\bf 2}; -1/6)\}$ \\

\hline

28&$\frac{3}{2} \cdot \textbf{{\color{blue}\{}}({\bf 6}, {\bf 1}; 4/3) , ({\overline{\bf 6}}, {\bf 1}; 2/3), ({\overline{\bf 1}}, {\bf 1}; -2)\textbf{{\color{blue}\}}}$ \\

&$-3 \cdot \textbf{{\color{blue}\{}}({\bf 3}, {\bf 1}; -1/3) , ({\overline{\bf 3}}, {\bf 1}; -1/3), ({\overline{\bf 6}}, {\bf 1}; 2/3)\textbf{{\color{blue}\}}}$ \\

\hline
\hline

\end{tabular}}
\caption{Identification of the T-II short-range eHSMs. For notations
  see Appendix and the main text.}
\label{t:HSMIdentTII}   
\end{table}   

%%
%\bibliography{references_0nubb.bib}
%%\bibliographystyle{h-physrev5}
%
%\end{document}

\end{document}